 \documentclass[twocolumn]{aastex62}
 \usepackage{subfigure}

\newcommand{\CI}{[C\,{\sc i}]}
\newcommand{\CII}{[C\,{\sc ii}]}
\newcommand{\Cone}{[C\,{\sc i}]\,(1$-$0)}
\newcommand{\Ctwo}{[C\,{\sc i}]\,(2$-$1)}
\newcommand{\COone}{CO\,(1$-$0)}
\newcommand{\RCI}{$R_\mathrm{[CI]}$}
\newcommand{\mum}{$\mu$m}

\shorttitle{Resolved carbon emission in nearby galaxies}
\shortauthors{Jiao et al.}

\begin{document}

\pdfoutput=1

\title{Resolved neutral carbon emission in nearby galaxies:\\
    \CI\ Lines as Total Molecular Gas Tracers}

\author{Qian Jiao}
\affil{Purple Mountain Observatory \& Key Lab of Radio Astronomy, Chinese Academy of Sciences (CAS), Nanjing 210008, China, jiaoqian@pmo.ac.cn, yugao@pmo.ac.cn}

\author{Yinghe Zhao}
\affiliation{Yunnan Observatories \& Key Laboratory for the Structure and Evolution of Celestial Objects, CAS, Kunming 650011, China, zhaoyinghe@ynao.ac.cn}
\affiliation{Center for Astronomical Mega-Science, CAS, 20A Datun Road, Chaoyang District, Beijing 100012, China}

\author{Nanyao Lu}
\affiliation{China-Chile Joint Center for Astronomy (CCJCA), Camino El Observatorio 1515, Las Condes, Santiago, Chile}
\affiliation{National Astronomical Observatories \& Key Lab of Radio Astronomy , CAS, Beijing 100012, China}

\author{Yu Gao}
\affiliation{Purple Mountain Observatory \& Key Lab of Radio Astronomy, Chinese Academy of Sciences (CAS), Nanjing 210008, China, jiaoqian@pmo.ac.cn, yugao@pmo.ac.cn}

\author{Dragan Salak}
\affiliation{School of Science and Technology, Kwansei Gakuin University, 2-1 Gakuen, Sanda, Hyogo 669-1337, Japan}

\author{Ming Zhu}
\affiliation{National Astronomical Observatories \& Key Lab of Radio Astronomy , CAS, Beijing 100012, China}

\author{Zhiyu Zhang}
\affiliation{Institute for Astronomy, University of Edinburgh, Royal Observatory, Blackford Hill, Edinburgh EH9 3HJ, UK}
\affiliation{ESO, Karl Schwarzschild Strasse 2, D-85748 Garching, Munich, Germany}

\author{Xuejian Jiang}
\affiliation{Purple Mountain Observatory \& Key Lab of Radio Astronomy, Chinese Academy of Sciences (CAS), Nanjing 210008, China, jiaoqian@pmo.ac.cn, yugao@pmo.ac.cn}

\author{Qinghua Tan}
\affiliation{Purple Mountain Observatory \& Key Lab of Radio Astronomy, Chinese Academy of Sciences (CAS), Nanjing 210008, China, jiaoqian@pmo.ac.cn, yugao@pmo.ac.cn}



\begin{abstract}

We present maps of atomic carbon \CI\,($^{3} \rm P_{1} \rightarrow {\rm ^3 P}_{0}$) and \CI\,($^{3} \rm P_{2} \rightarrow {\rm ^3 P}_{1}$) emission (hereafter \Cone\ and \Ctwo, respectively) at a linear resolution $\sim1\,$kpc scale for a sample of one H{\sc ii}, six LINER, three Seyfert and five starburst galaxies observed with the $Herschel\ Space\ Observatory$. We compare  spatial distributions of two \CI\ lines with that of CO $J=1\rightarrow 0$ (hereafter \COone) emission, and find that both \CI\ lines distribute similarly to \COone\ emission in most galaxies. We present luminosity ratio maps of $L'_\mathrm{[CI](1-0)}$/$L'_\mathrm{CO(1-0)}$, $L'_\mathrm{[CI](2-1)}$/$L'_\mathrm{CO(1-0)}$, $L'_\mathrm{[CI](2-1)}/L'_ \mathrm{[CI](1-0)}$ (hereafter $R_\mathrm{[CI]}$) and 70-to-160\,\mum\ far-infrared color of $f_{70}/f_{160}$. $L'_\mathrm{[CI](2-1)}$/$L'_\mathrm{CO(1-0)}$, $R_\mathrm{[CI]}$ and $f_{70}/f_{160}$ are centrally peaked in starbursts; whereas remain relatively constant in LINERs, indicating that star-forming activity can enhance carbon emission, especially for \Ctwo. We explore the correlations between the luminosities of \COone\ and \CI\ lines, and find that $L'_\mathrm{CO(1-0)}$ correlates tightly and almost linearly with both $L'_\mathrm{[CI](1-0)}$ and $L'_\mathrm{[CI](2-1)}$, suggesting that \CI\ lines, similar as \COone, can trace total molecular gas in H{\sc ii}, LINER, Seyfert and starburst galaxies on kpc scales. We investigate the dependence of $L'_\mathrm{[CI](1-0)}$/$L'_\mathrm{CO(1-0)}$, $L'_\mathrm{[CI](2-1)}$/$L'_\mathrm{CO(1-0)}$ and \CI\ excitation temperature of $T_\mathrm{ex}$ on dust temperature of $T_\mathrm{dust}$, and find non-correlation, a weak and modest correlation, respectively. The ratio of $L'_\mathrm{[CI](1-0)}$/$L'_\mathrm{CO(1-0)}$ stays smooth distribution in most galaxies, indicating that the conversion factor of \Cone\ luminosity to H$_2$ mass ($X_\mathrm{[CI](1-0)}$) changes with \COone\ conversion factor ($\alpha_\mathrm{CO}$) proportionally. Under optically thin and local thermodynamical equilibrium (LTE) assumptions, we derive a galaxy-wide average carbon excitation temperature of $T_\mathrm{ex} \sim 19.7 \pm 0.5\,$K and an average neutral carbon abundance of $X[\mathrm{CI}]/X[\mathrm{H_2}] \sim 2.5 \pm 1.0 \times 10^{-5}$ in our resolved sample, which is comparable to the usually adopted value of $3 \times 10^{-5}$, but $\sim3$ times lower than the carbon abundance in local (ultra-)luminous infrared galaxies ((U)LIRGs). We conclude that the carbon abundance varies in different galaxy types.

\end{abstract}


\keywords{ISM: atoms --- ISM: molecules --- ISM: abundances ---galaxies: starburst --- galaxies: spiral}

\section{Introduction} \label{sec:intro}

Carbon monoxide (CO) is widely used as molecular gas tracer \citep{Dickman et al. 1986, Sanders et al. 1991, Solomon et al. 2005, Bolatto et al. 2013}. However, several issues have limited the ability of CO in tracing molecular gas, such as the dependence of CO-to-H$_2$ conversion factor ($X_{\rm CO}$) on metallicity and gas density \citep[e.g.,][]{Downes Solomon 1998, Zhu et al. 2003, Leroy et al. 2011, Papadopoulos et al. 2012b, Bolatto et al. 2013}, and the impact from the cosmic microwave background (CMB) effects in the early universe \citep{Zhang et al. 2016}. 

Atomic carbon [C\,{\sc i}] $^{3} \rm P_{1} \rightarrow {\rm ^3 P}_{0}$ (rest frequency: 492.161 GHz, hereafter \Cone) and $^{3} \rm P_{2} \rightarrow {\rm ^3 P}_{1}$ (rest frequency: 809.344 GHz, hereafter \Ctwo) fine-structure transitions in its ground state received few attentions because \CI\ was pictured emanating only from a narrow \CII/\CI/CO transition zone according to  traditional photodissociation region (PDR) models \citep{Tielens et al. 1985, Hollenbach et al. 1991, Hollenbach et al. 1999}, and thus can not trace bulk molecular gas. However, more observations showed that \CI\ and CO coexist deep inside molecular clouds with a remarkably constant column density ratio of N(\CI)/N(CO)\citep[e.g.,][]{Ikeda et al. 1999, Ikeda et al. 2002, Ojha et al. 2001, Shimajiri et al. 2013}, rather than a thin layer on the surface of cloud, suggesting that the \CI\ lines might trace bulk molecular gas mass \citep{Papadopoulos et al. 2004a, Walter et al. 2011}. Furthermore, theoretical models show that \CI\ effectively traces molecular gas in solar metallicity clouds \citep[e.g.,][]{Offner et al. 2014, Glover et al. 2015}, and even remains a good molecular tracer in metal-poor \citep{Glover & Clark 2016} and high cosmic-ray environments where CO is severely depleted \citep[e.g.,][]{Bisbas et al. 2015, Bisbas et al. 2017, Papadopoulos et al. 2004a, Papadopoulos et al. 2018}. Assuming that \CI\ is well-mixed with H$_2$, \citet{Tomassetti et al. 2014} used a high-resolution hydrodynamic simulation and found that nearly all of the H$_2$ associated with the galaxy can be detected at redshifts z $<$ 4 through \CI\ lines with Atacama Large Millimeter Array (ALMA).

Recent observations demonstrate that \CI\ has remarkably good performance in tracing total molecular gas. The H$_2$ gas masses estimated with \Cone\ agree well with that derived from the standard method using CO in two typical ultra-luminous infrared galaxies (ULIRGs) NGC~6240 and Arp~220 \citep{Papadopoulos et al. 2004b}. \citet{Alaghband-Zadeh et al. 2013} found excellent agreement between the H$_2$ gas masses determined from \Cone\ and CO for their z$\sim 2.5$ submillimeter galaxies (SMGs); \citet{Weiss et al. 2003} derived similar total gas mass using \Cone\ and CO in Cloverleaf quasar at redshift of 2.5. \citet{Emonts et al. 2018} showed that \CI\ can be used to trace circumgalactic medium (CGM) of galaxies in the merging proto-custer, Spiderweb, at z = 2.2.  Additionally, \CI\ has already been used to trace molecular gas in distant starburst galaxies at z$\sim 4$ \citep{Bothwell et al. 2017}. 
 
Moreover, statistic studies also confirm that \CI\ lines are good indicators of total molecular gas mass for a large sample of local (U)LIRGs \citep{Jiao et al. 2017}, high-$z$ SMGs \citep{Yang et al. (2017)} and main-sequence galaxies \citep{Valentino et al. 2018}. As shown in \citet{Jiao et al. 2017}, the \COone\ luminosity is correlated linearly with both \CI\ luminosities for a sample of (U)LIRG observed with the $Herschel\ Space\ Observatory$ ($Herschel$; \citealt{Pilbratt et al. 2010}) Spectral and Photometric Imaging Receiver Fourier Transform Spectrometer(SPIRE/FTS; \citealt{Griffin et al. 2010, Swinyard et al. 2014}). The linear correlations indicate that both of the \CI\ lines can trace total molecular gas at least in (U)LIRGs. Furthermore, the tight correlation of CO and \Ctwo\ extends to high-redshift ($z$$\sim$$2-4$) SMGs \citep{Yang et al. (2017)}. 

However, most of these results are global characteristics averaged across whole galaxies. Further understanding properties of the \CI\ emission needs observations with higher spatial resolution and sensitivity. Before the advent of {\it Herschel}, few spatially resolved extragalactic \CI\ maps, generally only covering the central region, were available due to the low atmospheric transmission at \CI\ rest-frame frequencies. 

\citet{White et al. 1994} mapped the \CI\ and $\mathrm{CO\,(4-3)}$ emissions over the central $50\arcsec \times 30\arcsec$ of M~82, and found that \CI\ and CO are well mixed and have similar spatial distributions in the central region. \citet{Zhang et al. 2014} found similar results in the central region of nearby Seyfert galaxy, Circinus, and concluded that H$_2$ gas derived from \Cone\ is consistent with dust and multiple line CO modeling.  \citet{Krips et al. 2016} reported the first well-resolved interferometric (angular resolution $\sim 3''$) \CI\ map of the extragalactic source NGC~253, and found similar distributions between \CI\ and CO. In some central outflows, the \CI\ line has also been studied in detail with ALMA data at high resolutions (angular resolution $<1''$) \citep[e.g.,][]{Cicone et al. 2018, Miyamoto et al. 2018}.

The recent $Herschel$ space mission has produced resolved maps of nearby galaxies on sub-kpc scales. Here we present the $Herschel$ maps of the \Cone\ and \Ctwo\ lines toward 15 local spiral galaxies and analyze the properties of \CI\ emissions. We give a brief introduction about the observations, data reduction and analysis of the sample in section 2. In section 3 we present the results and discussion. In the last section we summarize the main conclusions. Throughout the paper, we use a Hubble constant of $H_0 = 70\mathrm{\ km\ s^{-1}\ Mpc^{-1}}$, $\Omega_\mathrm{M}=0.3$ and $\Omega_\mathrm{\lambda}=0.7$.

\section{Sample, data reduction and analysis}

\subsection{Sample}

\begin{table*}
\centering
\caption{The basic galaxy and observation information of the sample}
\label{The details of the sample}
\begin{tabular}{lccccccccccc}
	\hline
	\hline
	Galaxy\tablenotemark{a} & R.A. & Dec. & D & Ref.D\tablenotemark{b} & Proposal & Obs ID  & Type & CO telescope\tablenotemark{c} & Ref.CO\tablenotemark{d}   \\
	Name   & (hh$:$mm$:$ss) & (dd$:$mm$:$ss) &   (Mpc) & \\
	\hline
	M 51     & 13$:$29$:$52.7 & +47$:$11$:$42.6  & 8.2  & Da17 & KPGT\_cwilso01\_1 & 1342201202 & Seyfert & NRO-45m & K07  \\
	M~82 & 09$:$55$:$52.7 & +69$:$40$:$45.8  & 3.5 & Da17 & KPGT\_cwilso01\_1 & 1342208388  & starburst & NRO-45m  & S13  \\ 
	M~83     & 13$:$37$:$00.9 & $-$29$:$51$:$55.5 & 4.7  & Ra11 & KPGT\_cwilso01\_1 & 1342212345 & starburst & NRO-45m & K07  \\	 
	NGC~253  & 00$:$47$:$33.1 & $-$25$:$17$:$17.6 & 3.2  & Ra11 & KPGT\_rguesten\_1 & 1342210846 & starburst & NRO-45m & K07  \\ 
	NGC~891  & 02$:$22$:$33.4 & +42$:$20$:$56.9 & 10.2  & Ra11 & KPGT\_cwilso01\_1 & 1342213376  & H{\sc ii} & NRO-45m & S19  \\ 
	NGC~1068 & 02$:$42$:$40.7 & $-$00$:$00$:$47.8 &  14.4 & K07 & KPGT\_cwilso01\_1 & 1342213444 & Seyfert & NRO-45m & K07  \\ 
	NGC~3521  &  11$:$05$:$48.6 & $-$00$:$02$:$09.1 & 11.2 & Da17 & OT1\_jsmith01\_1 & 1342247743  & LINER & NRO-45m &  K07  \\ 
	NGC~3627 & 11$:$20$:$15.0 & +12$:$59$:$29.5  & 9.4 & Da17 & OT1\_jsmith01\_1 & 1342247604  & LINER & NRO-45m &  K07  \\ 
	NGC~4254 & 12$:$18$:$49.6 & +14$:$24$:$59.4 & 14.4 & Da17 & OT1\_jsmith01\_1 & 1342236997  & LINER & NRO-45m &  K07  \\ 
	NGC~4321 & 12$:$22$:$54.8 & +15$:$49$:$18.5  & 14.3 & Da17 & OT1\_jsmith01\_1 & 1342233784 & starburst & NRO-45m & K07  \\ 
	NGC~4569 & 12$:$36$:$49.8 & +13$:$09$:$46.6 & 9.9 & Da17 & OT1\_jsmith01\_1  & 1342248251 & Seyfert & NRO-45m & K07  \\ 
	NGC~4736 & 12$:$50$:$53.1 & +41$:$07$:$13.7  & 4.7  & Da17 & OT1\_jsmith01\_1 & 1342245851  & LINER & NRO-45m &  K07  \\ 	
	NGC~5055 & 13$:$15$:$49.3 & +42$:$01$:$45.4 & 7.9 & Da17 & OT1\_jsmith01\_1 & 1342237026 & LINER  & NRO-45m & K07  \\ 
	NGC~6946 & 20$:$34$:$52.3 & +60$:$09$:$14.1 & 6.8  & Da17 & OT1\_jsmith01\_1 & 1342243603 & starburst & NRO-45m & K07  \\  
	NGC~7331 & 22$:$37$:$04.0 & +34$:$24$:$55.9 & 14.5  & Da17 & OT1\_jsmith01\_1 & 1342245871 & LINER  & NRO-45m & S19 \\ 
	NGC~1482 * & 03$:$54$:$39.0 & $-$20$:$30$:$09.7  & 22.6  & Da17 & OT1\_jsmith01\_1 & 1342248233 & H{\sc ii} &  NRO-45m & S19 \\
	NGC~2976 * & 09$:$47$:$15.5 & +67$:$54$:$59.0 & 3.6 & Da17 & OT1\_jsmith01\_1 & 1342228706 & H{\sc ii} &  NRO-45m & S19 \\
	NGC~3077 * & 10$:$03$:$19.1 & +68$:$44$:$02.1 & 3.8  & Da17 & OT1\_jsmith01\_1 & 1342228745 & H{\sc ii} &  NRO-45m & S19 \\
	NGC~3351 * & 10$:$43$:$57.7 & +11$:$42$:$13.7 & 9.3  & Da17 & OT1\_jsmith01\_1 & 1342247117 &  starburst & NRO-45m & K07  \\
	NGC~4536 * & 12$:$34$:$27.1 & +02$:$11$:$17.3 & 14.5  & Da17 & OT1\_jsmith01\_1 & 1342237025 &  starburst & NRO-45m & K07  \\
	NGC~5457 * & 14$:$03$:$12.5 & +54$:$20$:$56.2 & 6.7  & Da17 & OT1\_jsmith01\_1 & 1342230417& H{\sc ii} & NRO-45m & K07 \\
	NGC~5713 * & 14$:$40$:$11.5 & $-$00$:$17$:$20.3 & 21.4  & Da17 & OT1\_jsmith01\_1 & 1342248250 & H{\sc ii} &  NRO-45m & S19 \\

	\hline
\end{tabular}
\tablenotetext{a}{The symbol ``*'' represents the 7 galaxies that have few \CI\ and/or \COone\ detections and are not included in the paper.}
\tablenotetext{b}{Reference to the distance of each galaxy. Da17 = \citealt{Dale et al. 2017}, Ra11 = \citealt{Radburn-Smith et al. 2011}, K07 = \citealt{Kuno et al. 2007}.}
\tablenotetext{c}{The observation telescope of \COone\ data. NRO-45m = Nobeyama 45-m telescope.}	
\tablenotetext{d}{Reference to the \COone\ data. K07 = \citealt{Kuno et al. 2007}, S13 = \citealt{Salak et al. 2013}, S19 = Sorai, K., et al., in preparation.}
\end{table*}

Our sample is primarily selected from the cross matching of the sample in \citet{Kamenetzky et al. 2016} (including all available extragalactic \CI\ observations in the $Herschel$ SPIRE/FTS Archive) with \COone\ samples in \citet{Kuno et al. 2007} \footnote{The \COone\ from \citet{Kuno et al. 2007} is published and available in the website: \url{http://www.nro.nao.ac.jp/~nro45mrt/html/COatlas/}}, \citet{Salak et al. 2013}, and Sorai, K., et al. (in preparation)\footnote{The \COone\  from Sorai, K., et al. (in preparation) is available in the website: \url{https://astro3.sci.hokudai.ac.jp/~radio/coming/data/}} observed with the Nobeyama 45-m telescope. We obtain 22 galaxies and list them in Table\,\ref{sec:intro}. Excluding 7 galaxies (with symbol ``*'' in Table\,\ref{sec:intro}) which have no more than 3 detected points of \CI\ and/or \COone, we focus on the remaining 15 galaxies. These 7 galaxies will be presented in the future paper that includes upper limits of current 15 galaxies. The final sample contains one H{\sc ii} galaxy (NGC~891), six LINER galaxies (NGC~3521, NGC~3627, NGC~4254, NGC~4736, NGC~5055, NGC~7331), three Seyfert galaxies (M~51, NGC~1068, NGC~4569), and five starburst galaxies (M~82, M~83, NGC~253, NGC~4321, NGC~6946).

The obtained \CI\ data of the final sample are observed by the $Herschel$ SPIRE/FTS from the following three projects: ``Beyond the Peak: Resolved Far-Infrared Spectral Mapping of Nearby Galaxies with SPIRE/FTS" (OT1\_jsmith01\_1, PI: J. D. Smith, e.g., \citealt[][]{Kamenetzky et al. 2016, Zhao et al. 2016}), ``Physical Processes in the Interstellar Medium of Very Nearby Galaxies" (KPGT\_cwilso01\_1, PI: C. D. Wilson, e.g., \citealt[][]{Spinoglio et al. 2012, Hughes et al. 2015}) and ``Physical and Chemical Conditions of the ISM in Galactic Nuclei" (KPGT\_rguesten\_1, PI: R. G\"usten, e.g., \citealt{Perez-Beaupuits et al. 2018}). 

\subsection{Data reduction}

\begin{table*}
\centering
\caption{The resolution, SFR, $\alpha_\mathrm{CO}$, $T_\mathrm{ex}$ and carbon abundance of each sample}
\label{sample}
\begin{tabular}{lccccccccc}
	\hline
	\hline
	Galaxy & \Cone  & \Ctwo & $\log(\sum \mathrm{SFR})$\tablenotemark{a} & $\alpha_\mathrm{CO}$\tablenotemark{b}   & $T_\mathrm{ex}$\tablenotemark{c} & $X\mathrm{[CI]_{10}}/X\mathrm{(H_2)}$\tablenotemark{d} & $X\mathrm{[CI]_{21}}/X\mathrm{(H_2)}$\tablenotemark{e} \\
	  & spatial scale & spatial scale & & & \\
	Name   & (kpc) & (kpc) & ($M_\odot\,\mathrm{yr^{-1}\,kpc^{-2}}$)  & $M_\odot\  (\mathrm{K\,km\,s^{-1}\,pc^2})^{-1}$ & (K) & $(10^{-5})$ & $(10^{-5})$ \\
	\hline
	M 51     & 1.5 & 1.4 &  $-2.2 \pm 0.1$ & -  & 19.3 $\pm$ 0.5 &  - & - \\ 
	M~82     & 0.7 & 0.6 & $-0.5 \pm 0.1$ & 1.0 (W01) & 25.6 $\pm$ 0.4 & 2.2 $\pm$ 0.8 & 2.2 $\pm$ 0.8 \\
	M~83     & 0.9 & 0.8 & - & - & 20.7 $\pm$ 0.3 & - & - \\ 
	NGC~253  & 0.6 & 0.6 & - & - & 23.7 $\pm$ 0.7 &  - & - \\ 
	NGC~891  & 1.9 & 1.8 & - & 2.2 (G93) & 17.4 $\pm$ 0.6 & 2.8 $\pm$ 1.0 & 2.8 $\pm$ 1.1 \\
	NGC~1068 & 2.7 & 2.6 & - &  - & 24.5 $\pm$ 0.4 &  - & -\\ 
	NGC~3521 & 2.1 & 2.0 & $-2.6 \pm 0.1$ & 7.3 (Sa13) & 17.8 $\pm$ 0.3 & 2.1  $\pm$ 0.8 & 2.2 $\pm$ 0.8 \\
	NGC~3627 & 1.8 & 1.6 & $-2.4 \pm 0.1$ & 1.8 (Sa13) & 20.1 $\pm$ 0.3 & 1.3 $\pm$ 0.5 & 1.3 $\pm$ 0.5 \\
	NGC~4254 & 2.7 & 2.6 & $-2.1 \pm 0.1$ & 4.7 (Sa13) & 14.1 $\pm$ 0.3 & 4.5 $\pm$ 1.7 & 4.6 $\pm$ 1.7 \\
	NGC~4321 & 2.7 & 2.6 & $-2.3 \pm 0.1$ & 2.2 (Sa13) & 17.2 $\pm$ 0.3 & 2.8 $\pm$ 1.0 & 2.8 $\pm$ 1.0 \\
	NGC~4569 & 1.9 & 1.7 & $-3.3 \pm 0.1$ & 1.2 (Sa13) & 20.4 $\pm$ 0.6 & 1.8 $\pm$ 0.7 & 1.8 $\pm$ 0.7 \\
	NGC~4736 & 0.9 & 0.8  & $-2.4 \pm 0.1$ & 1.1 (Sa13) & 22.9 $\pm$ 0.6 & 2.3 $\pm$ 0.8 & 2.3 $\pm$ 0.9 \\	
	NGC~5055 & 1.5 & 1.4 &  $-2.4 \pm 0.1$ & 4.0 (Sa13) & 16.2 $\pm$ 0.6 & 3.0 $\pm$ 1.1 & 3.0 $\pm$ 1.2\\
	NGC~6946 & 1.3 & 1.2  & $-1.8 \pm 0.1$ & 1.8 (Sa13) & 18.9 $\pm$ 0.3 & 1.8 $\pm$ 0.6 & 1.8 $\pm$ 0.7 \\ 
	NGC~7331 & 2.7 & 2.6 & $-2.7 \pm 0.1$ & 10.7 (Sa13) & 16.0 $\pm$ 0.3 & 3.1 $\pm$ 1.1 & 3.2 $\pm$ 1.2 \\

	\hline
\end{tabular}

\tablenotetext{a}{The star formation rate surface density of each galaxy from \citet{Calzetti et al. 2010}.}
\tablenotetext{b}{The CO-to-H$_2$ conversion factor ($\alpha_\mathrm{CO}$) and reference to the $\alpha_\mathrm{CO}$ of each galaxy. W01 = \citealt{Weiss et al. 2001}, G93 = \citealt{Guelin et al. 1993}, Sa13 = \citealt{Sandstrom et al. 2013}.}
\tablenotetext{c}{The galaxy average \CI\ excitation temperature.}
\tablenotetext{d}{The galaxy average carbon abundance calculated based on \Cone\ luminosity (details see section\ref{carbon-abundance}).}
\tablenotetext{e}{The galaxy average carbon abundance calculated based on \Ctwo\ luminosity (details see section\ref{carbon-abundance}).}
\end{table*}

The \CI\ data are reduced by the standard SPIRE/FTS reduction and calibration pipeline for mapping mode included in the Herschel Interactive Processing Environment (HIPE; \citealt{Ott 2010}) version 14.1. The fluxes of \CI\ lines are estimated by fitting the observed line profiles with the instrumental Sinc function, as discussed in detail in \citet{Zhao et al. 2013, Zhao et al. 2016} and \citet{Lu et al. 2017}. The \COone\ lines were observed with the Nobeyama 45-m telescope, and were collected from \citet{Kuno et al. 2007, Salak et al. 2013}; and Sorai, K., et al. (in preparation). We also retrieve the calibrated $Herschel$ Photodetector Array Camera and Spectrometer (PACS, \citealt{Poglitsch et al. 2010}) 70$\mu$m and 160$\mu$m photometric data, and 2MASS (short for The Two Micron All-Sky Survey) J-band images from the NASA/IPAC Infrared Science Archive. Table\,\ref{sec:intro} lists the basic galaxy and observation information for all galaxies.

The SPIRE/FTS beams can be approximated as Gaussian profiles with FWHMs of 38.6$"$ and 36.2$"$ at 492 GHz (the rest frequency of \Cone) and 809 GHz (the rest frequency of \Ctwo) \citep{Makiwa et al. 2013}, respectively. In terms of the integrated beam size, Gaussian beam approximations may differ from the true beams by no more than 20\% for \Cone\ and \Ctwo\ \citep{Makiwa et al. 2013}. The beams of the mapped \COone\ emission are 22$"$ for M~82 and 15$"$ for the other galaxies. In order to match the resolutions of \COone\ and \CI\ data, we convolve \COone\ images with convolution kernels generated by comparing the Nobeyama profile with SPIRE/FTS Gaussian profile of FWHM of 38.6$"$ and 36.2$"$ \citep{Aniano et al. 2011}. We also convolve the PACS 70$\mu$m and 160$\mu$m data to the lowest resolution (FWHM of 38.6$"$) using the same method. The \COone, PACS 70$\mu$m and 160$\mu$m data are finally regrided to the same pixel size as \CI\ data.

\CI\ fluxes are converted from the unit of $\mathrm{W\,m^{-2}}$ to $\mathrm{Jy\,km\,s^{-1}}$ using $F (\mathrm{W\,m^{-2}}) = \int F_\nu dv\ (\mathrm{Jy\,km\,s^{-1}}\times 3.3 \times 10^{-23}\nu (\mathrm{GHz})$ \citep{Kamenetzky et al. 2012}. \COone\ flux is converted from $\mathrm{K\,km\,s^{-1}}$ to $\mathrm{Jy\,km\,s^{-1}}$ using $\mathrm{S_v(Jy\ beam^{-1})/T_{mb}(K) = 2.4}$ \citep{Papadopoulos et al. 2012a}. The main beam efficiency is $\eta_{mb}=0.32$ for M~82 \citep{Salak et al. 2013}, and other galaxies already been converted to main beam temperatures \citep{Kuno et al. 2007}. Typical uncertainties of \Cone\ and \Ctwo\ line fluxes are 11\% and 13\%, respectively, which already include the absolute calibration uncertainty of 6\% for SPIRE FTS observations \citep{Swinyard et al. 2014}. The uncertainty of \COone\ flux is estimated to be $20\%$. Together with the convolution error which is less than $30\%$, the final uncertainty of \COone\ is $\sim 36\%$. 

We also compare our estimated \CI\ and \COone\ fluxes to those presented in \cite{Kamenetzky et al. 2016}, who estimated the line fluxes using a source/beam correction method as described in \citet{Kamenetzky et al. 2014, Kamenetzky et al. 2016}. Briefly, they firstly convolved the 250 $\mu$m image to a larger beam size $\Omega_b$, and measured the peak flux density. Then the ratio of this flux density to that of a $b=43''.5$ beam was used to scale the line flux measured from the SPIRE/FTS spectrum. This method has assumed that the line-to-continuum ratio is constant. However, \citet{Lu et al. 2017} have shown that the \Ctwo-to-continuum ratio is dependent on the far-infrared color. Indeed, the \Cone\ fluxes in \cite{Kamenetzky et al. 2016} are comparable to ours, while their \Ctwo\ fluxes are overall larger than our results. Therefore we exclude their sample from our following analysis.

\subsection{Analysis}

\begin{figure*}

\includegraphics[bb=60 168 556 725, clip]{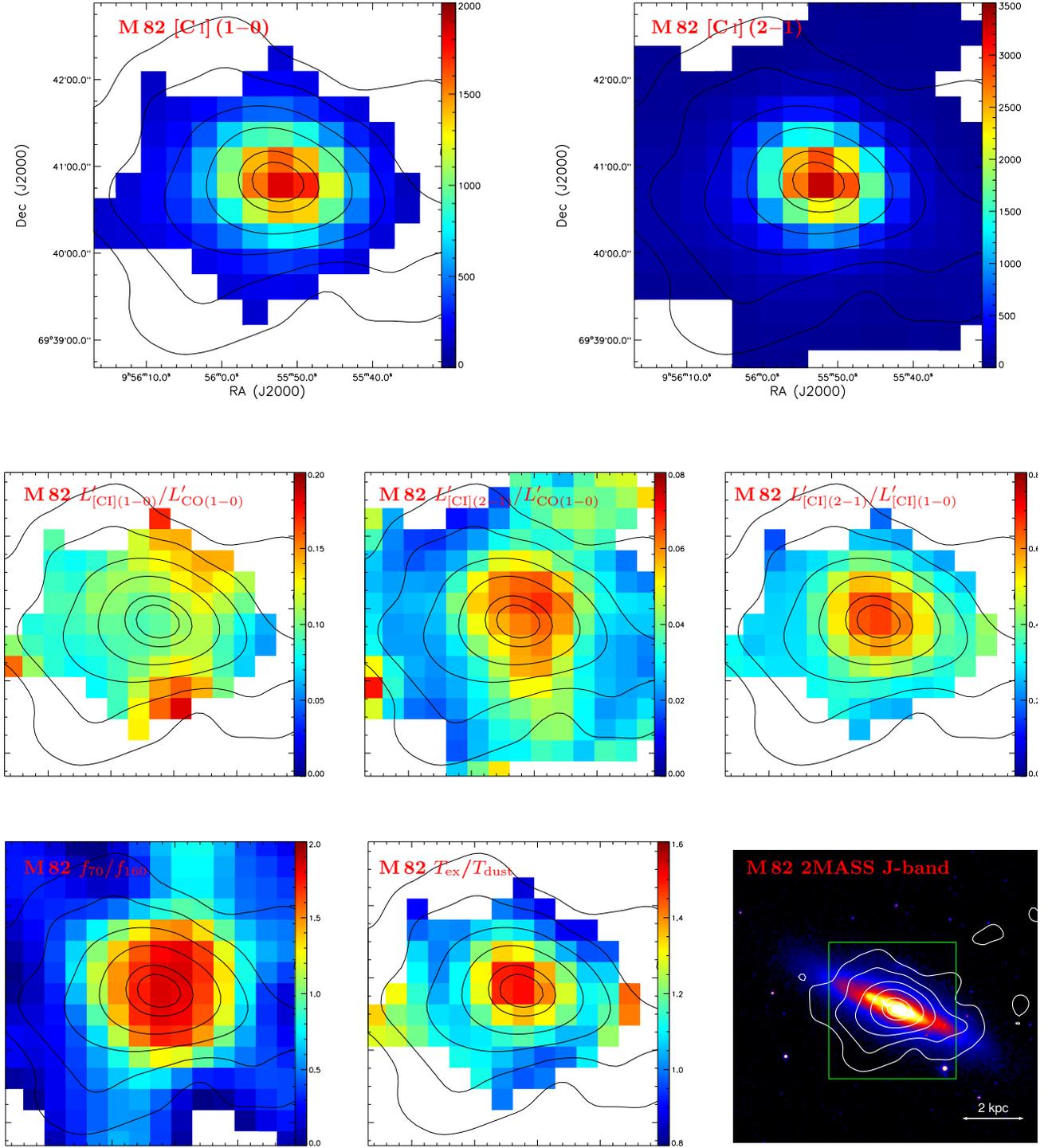}
\caption{Top row: the \Cone\ and \Ctwo\ integrated intensity (in unit of $\mathrm{Jy\,km\,s^{-1}}$) distributions in M~82; middle row: the \CI-CO luminosity ratios of $L'_\mathrm{[CI](1-0)}/L'_\mathrm{CO(1-0)}$ and $L'_\mathrm{[CI](2-1)}/L'_\mathrm{CO(1-0)}$, and the \CI\ luminosity ratio of $L'_\mathrm{[CI](2-1)}/L'_\mathrm{[CI](1-0)}$ (\RCI) distributions in M~82; bottom row: the color of $f_{70}/f_{160}$ distribution in M~82, and the ratio of \CI\ excitation temperature to dust temperature of $T_\mathrm{ex}$/$T_\mathrm{dust}$, and the 2MASS J-band photometry of M~82. The black and white contours (in panel 2MASS J-band) in these panels are the integrated intensity of the convolved \COone\ emission \citep{Salak et al. 2013} at 5, 10, 20, 40, 80, 100$\sigma$ levels with $\sigma$=3.1 $\mathrm{K\,km\,s^{-1}}$, respectively. The green box in panel 2MASS J-band shows the region of \CI\ observation with $Herschel$. The complete sample figures (15 galaxies) are shown in the figure Set, and the figure Set is available in the online journal.}
\label{fig:M82}
\end{figure*}

The sample covers several orders of magnitude in star formation rate (SFR, see Table \ref{sample}) and a variety of galaxy types. For each galaxy, we present both \Cone\ and \Ctwo\ emission maps (with a map cutoff at signal-to-noise (S/N) $=$ 3) in top rows of Figure\,\ref{fig:M82} and figure set. The complete figure set (15 galaxies) is available in the online version. The detected \Ctwo\ regions are larger than the \Cone\ regions in most cases due to the fact that the [CI](1-0) line is located near  the low frequency end of the FTS where noise is somewhat elevated \citep{Swinyard et al. 2014}. Spatial resolutions of the \CI\ lines for each galaxies are shown in Table\,\ref{sample}. Among our sample, M~82 owns the best resolved maps and resolutions. We will use M~82 as an example to analyze the properties of \CI\ emissions in section\,\ref{Distribution section}.

For the 15 spatially resolved galaxies, we calculate their \CI\ and \COone\ luminosities  using \citet{Papadopoulos et al. 2012a}:
\begin{equation}
	L'_x =  3.25 \times 10^3\,
	           \left [
	              \frac{D_\mathrm{L}^2 (\mathrm{Mpc})} {1+z}
	           \right ]
	           \left (
	               \frac{v_{x,\mathrm{rest}}} {100 GHz}
	           \right )^{-2}
	           \left [
	                \frac{\int_{\Delta \mathrm{V}} \, S_v \, d\mathrm{V}} 
	                       {Jy\, km\, s^{-1}}
	           \right ], 
\label{luminosity}
\end{equation}
where $L'_x$ is in unit of $\mathrm{K\, km\, s^{-1}}$, $v_{x,\mathrm{rest}}$ is the rest frequency and $S_v$ represents the line flux density.

The $L'_x$ ratio of two lines in the same source equals to the ratio of their intrinsic brightness temperatures \citep{Solomon et al. 2005}. We calculate the \CI-CO luminosity ratios of $L'_\mathrm{[CI](1-0)}/L'_\mathrm{CO(1-0)}$ and $L'_\mathrm{[CI](2-1)}/L'_\mathrm{CO(1-0)}$ after regridding the convolved \COone\ data to the same resolution and pixel size of \CI\ data, respectively. The \CI\ luminosity ratio, $R_\mathrm{[CI]} = L'_\mathrm{[CI](2-1)}/L'_ \mathrm{[CI](1-0)}$, is calculated directly for each map pixel with both \CI\ detections without beam matching since the FWHMs of \Cone\ and \Ctwo\ are similar. Maps of $L'_\mathrm{[CI](1-0)}/L'_\mathrm{CO(1-0)}$, $L'_\mathrm{[CI](2-1)}/L'_\mathrm{CO(1-0)}$, and \RCI\ are shown in the middle rows of Figure\,\ref{fig:M82} and figure set, respectively.

Far-infrared color $f_{70}/f_{160}$ can be used as an indicator of the intensity of the ambient UV field (thus the dust temperature). We also measure the $f_{70}/f_{160}$ for each galaxy with 70~$\mu$m and 160~$\mu$m fluxes convolved and regridded to the resolution and pixel scale of \Cone. Assuming a grey body with dust emissivity index $\beta$ = 2, we further calculate dust temperature of $T_\mathrm{dust}$ using the ratio of $f_{70}/f_{160}$. The \CI\ excitation temperature of $T_\mathrm{ex}$ can be estimated using \RCI\ (see details in Section\ref{carbon-abundance}). In order to compare $T_\mathrm{ex}$ to $T_\mathrm{dust}$, we also map the ratio of $T_\mathrm{ex}$/$T_\mathrm{dust}$ for each galaxy. The distributions of $f_{70}/f_{160}$ and $T_\mathrm{ex}$/$T_\mathrm{dust}$ are shown in bottom rows in Figure\,\ref{fig:M82} and figure set. For comparison, we also show the 2MASS J-band image in Figure\,\ref{fig:M82} and figure set for each galaxy.

\figsetstart
\figsetnum{1}
\figsettitle{Maps of the complete sample}

\figsetgrpstart
\figsetgrpnum{1.1}
\figsetgrptitle{Image of M82}
\figsetplot{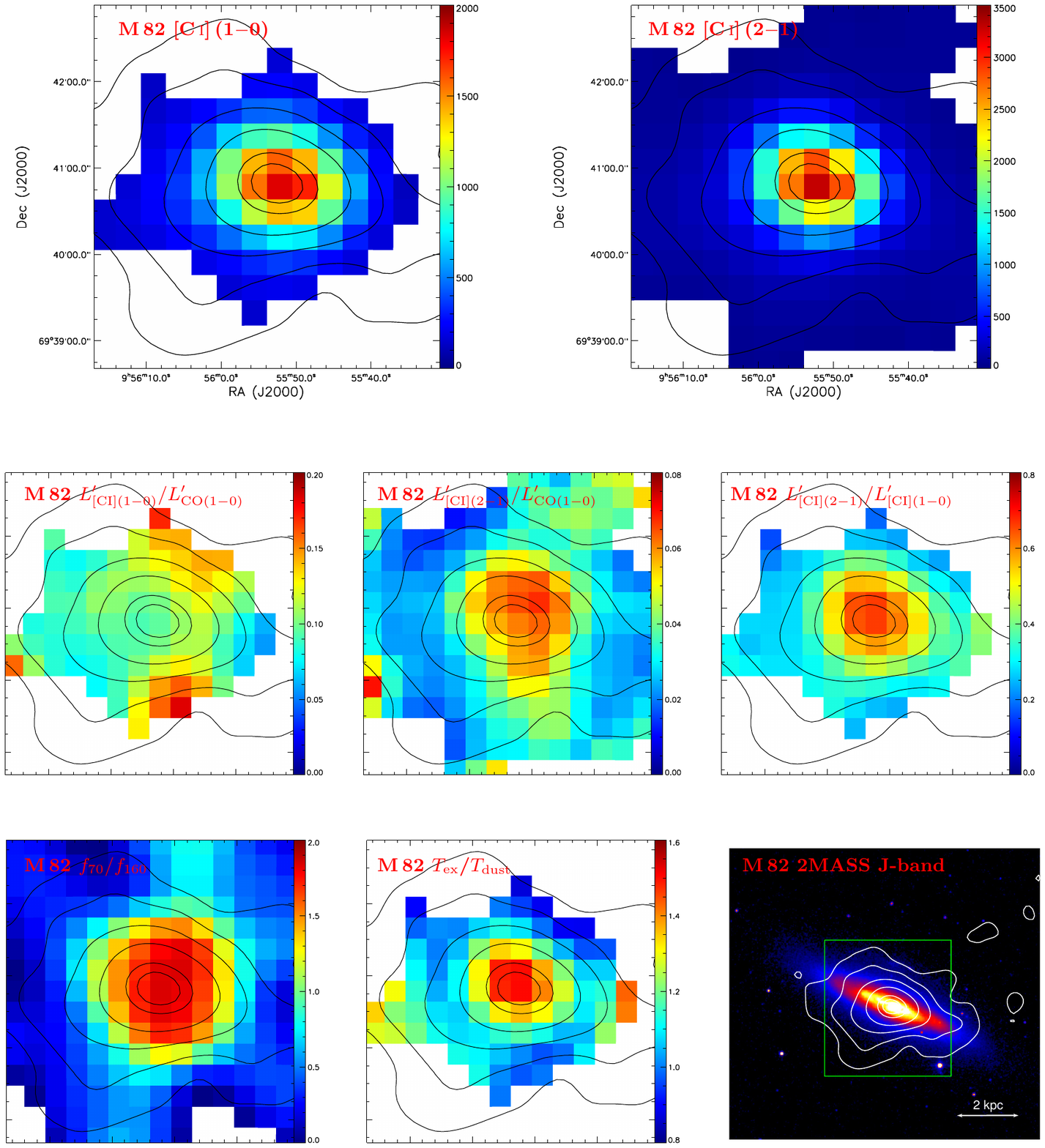}
\figsetgrpnote{Images collected by Chris Dolan.}
\figsetgrpend

\figsetgrpstart
\figsetgrpnum{1.2}
\figsetgrptitle{Image of M51}
\figsetplot{M51_figure.pdf}
\figsetgrpnote{----- Same as Figure\,\ref{fig:M82} but for M~51. The black and white contours are the integrated intensity of the convolved \COone\ emission \citep{Kuno et al. 2007} at 10, 40, 100, 200, 300, 400$\sigma$ levels with $\sigma$=0.5 $\mathrm{K\,km\,s^{-1}}$, respectively.}
\figsetgrpend

\figsetgrpstart
\figsetgrpnum{1.3}
\figsetgrptitle{Image of M83}
\figsetplot{M83_figure.pdf}
\figsetgrpnote{----- Continued but for M~83. The black and white contours are the integrated intensity of the convolved \COone\ emission \citep{Kuno et al. 2007} at 10, 20, 30, 50, 100, 140$\sigma$ levels with $\sigma$=1.4 $\mathrm{K\,km\,s^{-1}}$, respectively.}
\figsetgrpend

\figsetgrpstart
\figsetgrpnum{1.4}
\figsetgrptitle{Image of NGC253}
\figsetplot{NGC253_figure.pdf}
\figsetgrpnote{----- Continued but for NGC~253. The black and white contours are the integrated intensity of the convolved \COone\ emission \citep{Kuno et al. 2007} at 3, 10, 20, 30, 60, 120$\sigma$ levels with $\sigma$=6.6 $\mathrm{K\,km\,s^{-1}}$, respectively.}
\label{fig:NGC253}
\figsetgrpend

\figsetend

\section{Results and Discussion} 
\subsection{The distributions of \CI\ and \COone\ emission}
\label{Distribution section}

In the following, we briefly discuss galaxy M~82 before summarizing properties of the complete sample. M~82 (also known as NGC~3034) is one of the nearest (3.53\,Mpc, \citealt{Karachentsev et al. 2002}) galaxies, and famous for its intense starburst and prominent superwind associated with large-scale outflows \citep[e.g.,][]{Walter et al. 2002, Salak et al. 2013}. This galaxy has been well studied at numerous wavelengths. Radio \citep[e.g.,][]{Kronberg et al. 1985, Wills et al. 1999} and infrared \citep[e.g.,][]{Telesco & Gezari 1992} observations showed active star formation in the central region. The central star burst region contains bulk of molecular gas, and the starburst affects the physical conditions of the molecular gas \citep[e.g.,][]{Henkel & Bally 1985, Wild et al. 1992, Brouillet & Schilke 1993, Weiss et al. 2001, Salak et al. 2013}.

The top row of Figure\,\ref{fig:M82} shows the \Cone\ and \Ctwo\ integrated intensity distributions of M~82. As shown in Table\,\ref{sample}, the spatial resolutions of \Cone\ and \Ctwo\ maps are $\sim 0.7\ $kpc and $\sim0.6\ $kpc with each pixel size of $\sim 0.3\ $kpc, respectively. The middle row of Figure\,\ref{fig:M82} shows the luminosity ratios of $L'_\mathrm{[CI](1-0)}/L'_\mathrm{CO(1-0)}$, $L'_\mathrm{[CI](2-1)}/L'_\mathrm{CO(1-0)}$ and \RCI. The distributions of $f_{70}/f_{160}$ and $T_\mathrm{ex}$/$T_\mathrm{dust}$ of M~82 are shown in bottom row of Figure\,\ref{fig:M82}, respectively. We also present  the 2MASS J-band image of M~82 with the \CI\ observation region labeled as green box in bottom row of Figure\,\ref{fig:M82}. The black and white contours are the integrated intensity of convolved \COone\ emission (FWHM of 38.6$"$) \citep{Salak et al. 2013} at 5, 10, 20, 40, 80, 100$\sigma$ levels with $\sigma$=3.1 $\mathrm{K\,km\,s^{-1}}$, respectively.    

Both of the \CI\ lines in M~82 are enhanced in the central region and trail off into the outskirt, and have similar spatial distributions to \COone. More specially, $L'_\mathrm{[CI](1-0)}/L'_\mathrm{CO(1-0)}$ is $\sim$ $0.05-0.18$, and $L'_\mathrm{[CI](2-1)}/L'_\mathrm{CO(1-0)}$ is in the range of $\sim 0.01-0.07$ though it shows centrally-peaked distribution. These two images indicate that both of $L'_\mathrm{[CI](1-0)}/L'_\mathrm{CO(1-0)}$ and $L'_\mathrm{[CI](2-1)}/L'_\mathrm{CO(1-0)}$ ratios are nearly constant. \citet{White et al. 1994} derived an average abundance ratio of CI/CO $\sim 0.5$ across most of the M~82 nucleus. \citet{Fixsen et al. 1999} found a constant ratio of $L'_\mathrm{[CI](1-0)}/L'_\mathrm{CO(1-0)} = 0.15 \pm 0.1$ in the Milky way, and \citet{Gerin & Phillips 2000} reported $L'_\mathrm{[CI](1-0)}/L'_\mathrm{CO(1-0)} = 0.2 \pm 0.2$ in a sample of nearby spiral, irregular, interacting and merging galaxies. All of these results suggest that both \CI\ lines may have a good correlation with CO. In the next section, we will further analyze their correlations. 
    
$R_\mathrm{[CI]}$, $f_{70}/f_{160}$, and $T_\mathrm{ex}$/$T_\mathrm{dust}$ in Figure\,\ref{fig:M82} also show centrally-peaked distributions with $R_\mathrm{[CI]} \sim 0.6$ in the center and $R_\mathrm{[CI]} \sim 0.2-0.4$ in the outer region. In consideration of the similar distributions between $f_{70}/f_{160}$ and $R_\mathrm{[CI]}$, one can see obviously that higher temperature gives higher \CI\ excitation. This is consistent with our previous result in \citet{Jiao et al. 2017}. The \Ctwo\ excitation energy (64~K) is significantly higher than the typical \CI\ excitation temperature ($11-34\,$K, see details in Section\ref{carbon-abundance}) in our system, whereas the excitation energy of \Cone\ line (24~K) is more similar to the \CI\ excitation temperature. Consequently the \Ctwo\ line is more sensitive to the temperature. Additionally, according to the large-scale COBE maps, \citet{Bennett et al. 1993} reported $R_\mathrm{[CI]} = 0.35$ in the Galactic center, and $R_\mathrm{[CI]} = 0.17$ in the inner Galactic disk, which is smaller than that in M~82. Higher \RCI\ in M~82 compared to the Galaxy, as well as the enhancement of \RCI\ in the central regions of both M~82 and Milky Way, indicate that active star-formation may enhance carbon excitation \citep{Stutzki et al. 1997, Krips et al. 2016}.

The \CI\ lines in starburst galaxies M~83, NGC~253, NGC~4321 as well as NGC~6946 show central enhancement distributions, and have similar distributions to \COone, which are consistent with the result in M~82. Moreover, the $L'_\mathrm{[CI](2-1)}/L'_\mathrm{CO(1-0)}$, \RCI\ and $f_{70}/f_{160}$ in these starburst galaxies also show similar central enhanced  distribution as \CI\ lines, whereas relatively constant distributions are found in galaxies NGC~891, NGC~3521, NGC~3627, NGC~4254, and NGC~7331. This also indicates that the starburst can enhance carbon excitation. The Seyfert NGC~1068 is also classified as a powerful starburst galaxy with a SFR$=37\,M_\odot\,\mathrm{yr^{-1}}$ in the inner disk \citep{Planesas et al. 1989, Garcia-Burillo et al. 2014}, and also shows centrally enhanced $L'_\mathrm{[CI](2-1)}/L'_\mathrm{CO(1-0)}$, \RCI\ and $f_{70}/f_{160}$ emissions. The color of $f_{70}/f_{160}$ in NGC~4736 shows centrally-peaked distribution, and its \RCI\ is slightly higher than other LINER galaxies. This may due to the fact that the center region of NGC~4736 is likely a ``poststarburst$"$ phase \citep{Walker et al. 1988}. NGC~3521 exhibits a kpc-scale central depletion of molecular gas \citep{Helfer et al. 2003, Kuno et al. 2007, Nishiyama 2001}, which is also weekly shown in its convolved \COone\ flux contours. While both \CI\ lines of NGC~3521 show a peak in the nucleus, and trail off into the outer region. This may be related to the poor $Herschel/FTS$ resolutions at the \CI\ frequencies which can't resolve the center region well. Higher spatial resolution data are needed for further analysis.

\begin{figure}[!hptb]
  \centering
  \scriptsize      
\includegraphics[width=0.28\textwidth, angle=270, bb= 12 40 423 700]{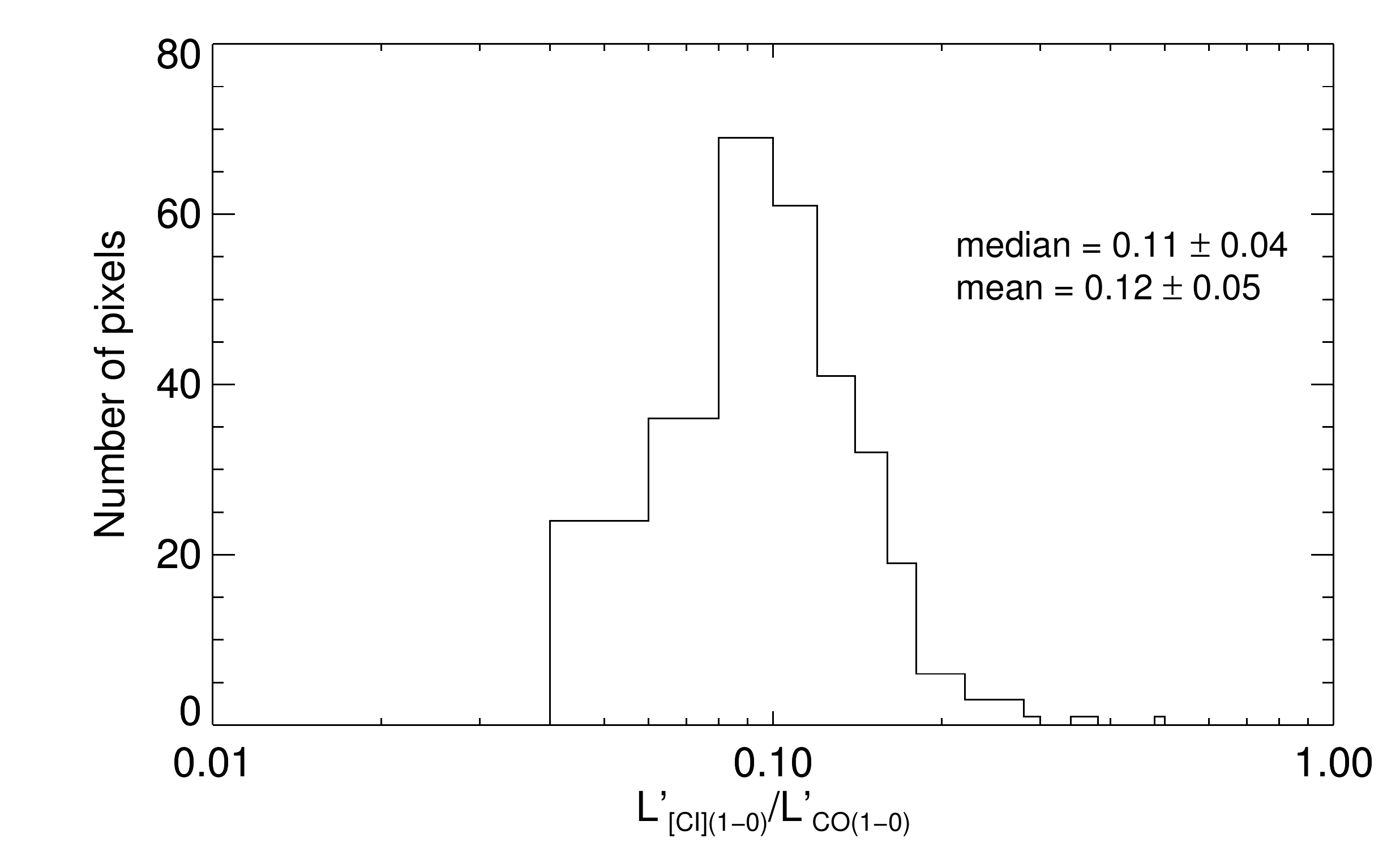} 
\includegraphics[width=0.28\textwidth, angle=270, bb= 12 40 423 700]{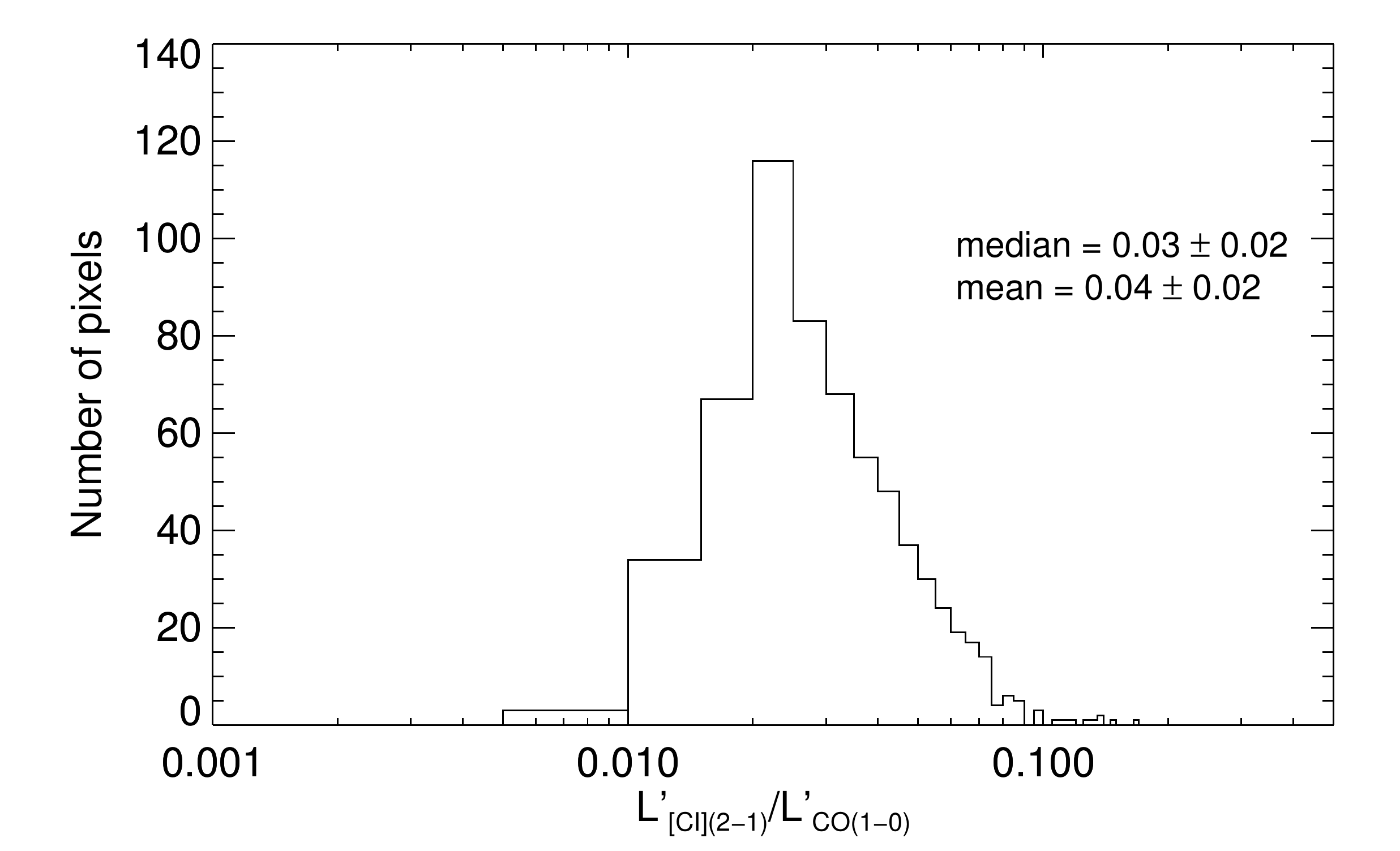} 
\includegraphics[width=0.28\textwidth, angle=270, bb= 12 40 423 700]{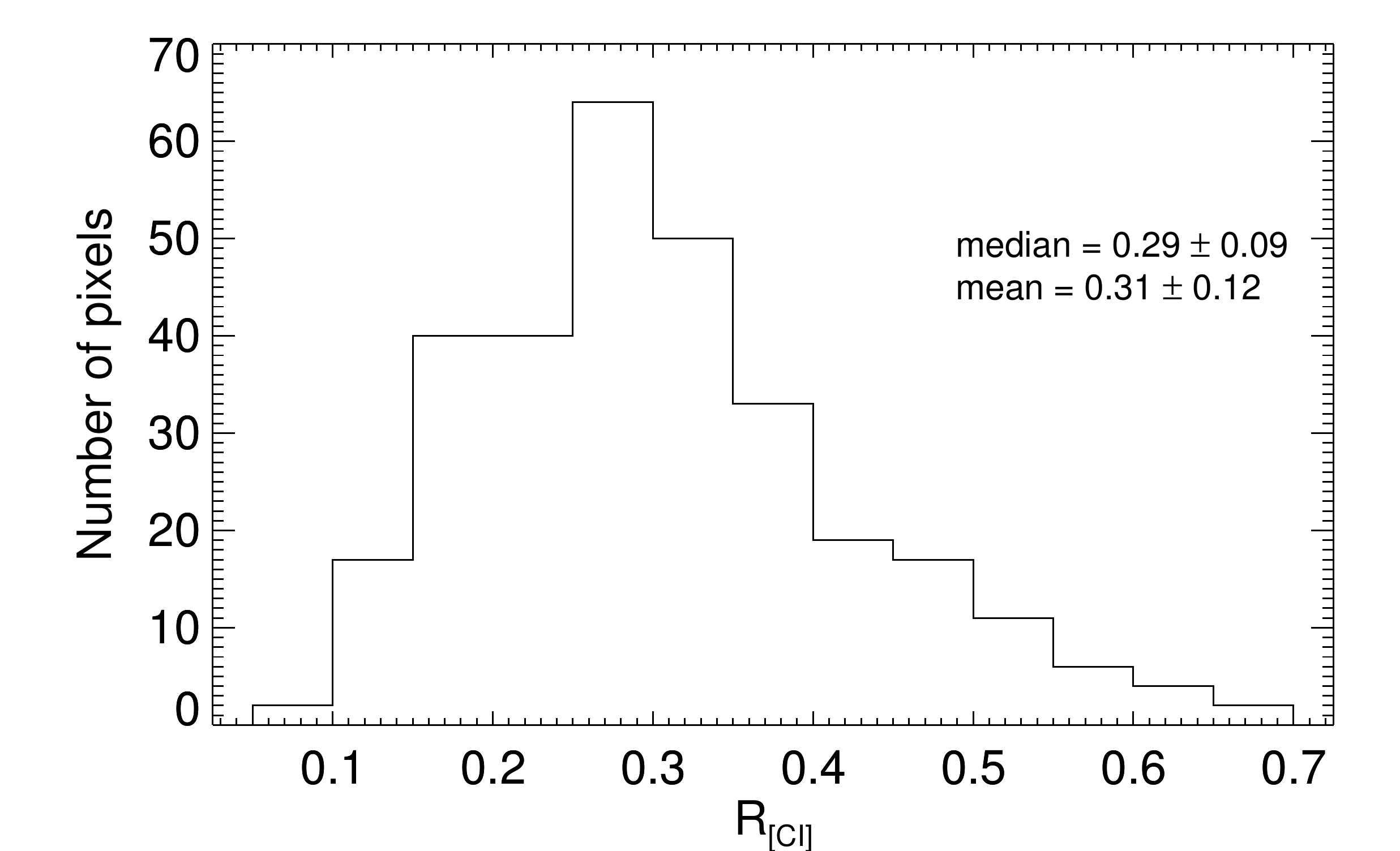}                           
\caption{Distributions of the $L'_\mathrm{[CI](1-0)}/L'_\mathrm{CO(1-0)}$ (top panel), $L'_\mathrm{[CI](2-1)}/L'_\mathrm{CO(1-0)}$ (middle panel), and \RCI\ (bottom panel) for the total resolved galaxies with median and mean values labelled on.}
\label{fig:Lci_Lco}
\end{figure}

In Figure\,\ref{fig:Lci_Lco}, we show the distributions of $L'_\mathrm{[CI](1-0)}/L'_\mathrm{CO(1-0)}$ (top panel), $L'_\mathrm{[CI](2-1)}/L'_\mathrm{CO(1-0)}$ (middle panel), and $R_\mathrm{{[CI]}}$ (bottom panel) for the total resolved galaxies with  median and mean values labelled on. The uncertainties in the median and mean values represent the median absolute deviations (MAD) and standard deviations of the resolved galaxies, respectively. The luminosity ratios of  $L'_\mathrm{[CI](1-0)}/L'_\mathrm{CO(1-0)}$ and $L'_\mathrm{[CI](2-1)}/L'_\mathrm{CO(1-0)}$ are constant for each galaxies, which is consistent with the results in Figure\,\ref{fig:M82} and figure set. In particular, \RCI\ doesn't vary too much for the sample with  median value of $R_\mathrm{{[CI]}}=0.29 \pm 0.09$ and mean value of $R_\mathrm{{[CI]}}=0.31 \pm 0.12$. This result is lower than the derived average \RCI\ $\sim 0.55 \pm 0.15$ in SMGs and quasar hot galaxies (QSOs) at high redshift (z$>$2) \citep{Walter et al. 2011}. Specially, \citet{Israel et al. 2015} also derived average line intensity ratios of \Ctwo/\Cone\ of $2.17 \pm 0.13$ for (U)LIRGs and $1.86 \pm 0.12$ for starbursts, which translate to luminosity ratios of $R_\mathrm{{[CI]}}=0.49 \pm 0.03$ and $R_\mathrm{{[CI]}}=0.42 \pm 0.03$, respectively. Compare to our resolved galaxies, the \RCI\ might be higher for (U)LIRGs and high-z galaxies which tend to be infrared warm objects.

\subsection{\CI\ lines as total molecular gas tracers}

\subsubsection{\CI-CO correlation}
\label{CI-CO correlation}

\begin{figure*}[!htpb]
   \centering
  \scriptsize      
\centerline{\includegraphics[width=0.78\textwidth, bb = 35 13 483 338]{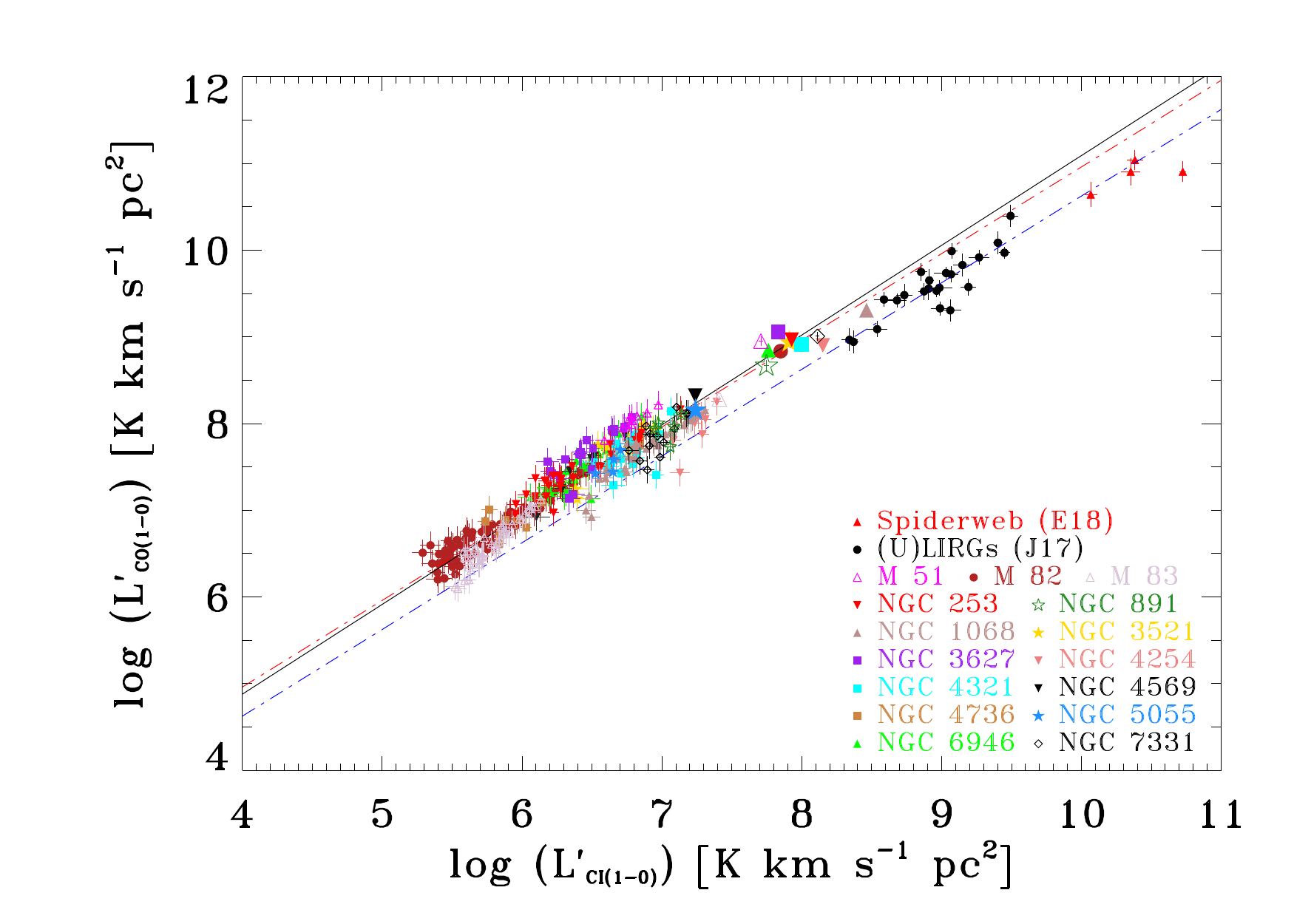}} 
\centerline{\includegraphics[width=0.78\textwidth, bb = 35 13 483 338]{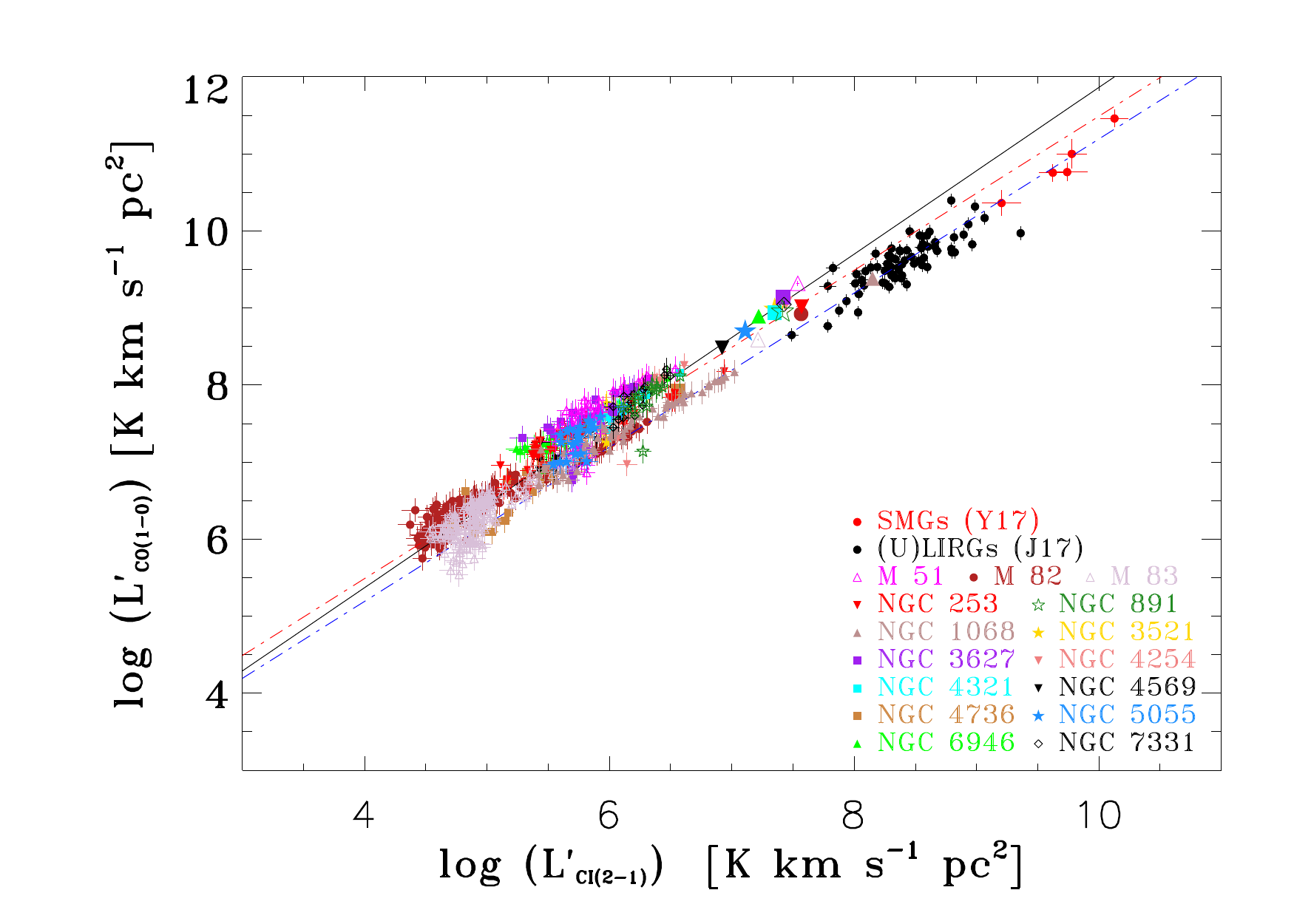}}                            

\caption{The luminosities of \COone\ are plotted against the luminosities of \Cone(top panel) and \Ctwo(bottom panel) for each resolved galaxies of M~51 (magenta triangles), M~82 (filled firebrick circles), M~83 (thistle triangles), NGC~253 (filled red inverted triangles), NGC~891 (forest green stars), NGC~1068 (filled rosy brown triangles), NGC~3521 (filled gold stars), NGC~3627 (filled purple squares), NGC~4254 (filled light coral inverted triangles), NGC~4321 (filled cyan squares), NGC~4569 (filled black inverted triangles), NGC~4736 (filled peru squares), NGC~5055 (filled dodger blue stars), NGC~6946 (filled green triangles), and NGC~7331 (black diamonds). The larger symbols are total values for each galaxies. For these resolved galaxies, the best-fit relations with a free slope are shown by the black lines, and best-fit relations with a fixed slope of 1 are shown by the dash-dot (red) lines. The filled black circles are (U)LIRGs from \citet{Jiao et al. 2017}. The filled red triangles are the central radio galaxy and its satellite galaxies, and the CGM of the massive Spiderweb Galaxy, a proto-cluster at z=2.2 from \citet{Emonts et al. 2018}. The filled red circles are high-redshift SMGs from \citet{Yang et al. (2017)}.   For the (U)LIRGs, Spiderweb Galaxy (for \Cone\ only) and SMGs (for \Ctwo\ only), the best-fit relations with a fixed slope of 1 are shown as the dash-dot (blue) lines.}
 
\label{fig:CIvsCO}
\end{figure*}

On galactic scales, \cite{Jiao et al. 2017} have shown that $L'_\mathrm{CO(1-0)}$ correlates almost linearly with both $L'_ \mathrm{[CI](1-0)}$ and $L'_\mathrm{[CI](2-1)}$ in (U)LIRGs, suggesting that the \CI\ lines can trace total molecular gas mass at least for (U)LIRGs. In order to further understand the \CI\ lines in galaxy, we need sub-galactic scale data and more galaxy types other than (U)LIRGs. The relatively constant ratios of $L'_\mathrm{[CI](1-0)}/L'_\mathrm{CO(1-0)}$ and $L'_\mathrm{[CI](2-1)}/L'_\mathrm{CO(1-0)}$ shown in Figure\,\ref{fig:Lci_Lco}, suggest that the relations of \CI\ and CO in (U)LIRGs \citep{Jiao et al. 2017} may could extend to sub-kpc scales of H{\sc ii}, LINREs, Seyfert galaxies and starbursts. In this section, we use resolved sample to investigate the properties of \CI\ lines as molecular tracers.

Figure\,\ref{fig:CIvsCO} shows the correlations of $L'_\mathrm{CO(1-0)}$$-$$L'_\mathrm{[CI](1-0)}$ (top panel) and $L'_\mathrm{CO(1-0)}$$-$$L'_\mathrm{[CI](2-1)}$ (bottom panel) for each pixels with detected \CI\ emission in our resolved galaxies: M~51 (magenta triangles), M~82 (filled firebrick circles), M~83 (thistle triangles), NGC~253 (filled red inverted triangles), NGC~891 (forest green stars), NGC~1068 (filled rosy brown triangles), NGC~3521 (filled gold stars), NGC~3627 (filled purple squares), NGC~4254 (filled light coral inverted triangles), NGC~4321 (filled cyan squares), NGC~4569 (filled black inverted triangles), NGC~4736 (filled peru squares), NGC~5055 (filled dodger blue stars), NGC~6946 (filled green triangles), and NGC~7331 (black diamonds). Moreover, we also show the location of a galaxy using the integrated line emission by a larger-sized symbol of the same shape and color as its sub-galactic data points. For comparison, we also present the (U)LIRGs from \citet{Jiao et al. 2017} as filled black circles, the central radio galaxy and its satellite galaxies, and the CGM of the massive Spiderweb Galaxy, a proto-cluster at z=2.2 from \citet{Emonts et al. 2018} as filled red triangles, and lensing corrected SMGs from \citet{Yang et al. (2017)} as filled red circles in Figure\,\ref{fig:CIvsCO}. Point (the rightmost filled red triangle in the top panel of Figure\,\ref{fig:CIvsCO}) with a high $L'_\mathrm{[CI](1-0)}/L'_\mathrm{CO(1-0)}$ $\sim$0.67 is the central radio galaxy MRC 1138-262 which may has cloud-heating mechanisms due to cosmic rays, jet-induced shocks, or gas turbulence \citep{Emonts et al. 2018}.

The two panels in Figure\,\ref{fig:CIvsCO} show that the $L'_\mathrm{CO(1-0)}$ is well correlated with both $L'_\mathrm{[CI](1-0)}$ and $L'_\mathrm{[CI](2-1)}$ even at sub-galactic scales with corresponding correlation coefficients of 0.95 and 0.94 for these resolved galaxies. More specifically, we fit these resolved galaxies using unweighted linear least-squares with a geometrical mean functional relationship \citep{Isobe et al. 1990}, which gives: 

\begin{equation}
\log L'_\mathrm{CO(1-0)} = (0.74 \pm 0.12) + (1.04\pm 0.02) \ \log L'_\mathrm{[CI](1-0)},
\label{COvsCI10}
\end{equation}
and
\begin{equation}
\log L'_\mathrm{CO(1-0)} =  (1.04 \pm 0.08) + (1.08 \pm 0.02) \ \log L'_\mathrm{[CI](2-1)},
\label{COvsCI21}
\end{equation}
with vertical scatters of 0.17 dex and 0.22 dex, respectively. The fitted results are shown in Figure\,\ref{fig:CIvsCO} as black lines. These tight and nearly linear relations imply that the \COone\ might arises from similar regions with \CI\ lines even at sub-galactic scale of $\sim1$ kpc size, which is well agree with the morphological results shown in Figure\,\ref{fig:M82}. The nearly linear correlations also indicate that the \CI\ lines, similar as \COone, can be used to trace the bulk of H$_2$ gas mass in H{\sc ii}, LINER, Seyfert and starburst galaxies on kpc scales.

In order to minus any systemic uncertainties, and obtain the linear correlations for $L'_\mathrm{CO(1-0)}$$-$$L'_\mathrm{[CI]}$ as well, we also fitted their relations with a fixed slope of 1 for these resolved galaxies, which gives:  
 \begin{equation}
\log L'_\mathrm{CO(1-0)} = (0.96 \pm 0.01) +   \log L'_\mathrm{[CI](1-0)},
\label{COvsCI10mcfit}
\end{equation}
and 
\begin{equation}
\log L'_\mathrm{CO(1-0)} =  (1.49 \pm 0.01) +  \log L'_\mathrm{[CI](2-1)},
\label{COvsCI21mcfit}
\end{equation}
with scatters of 0.17 dex and 0.22 dex, respectively. The fitted results are plotted in Figure\,\ref{fig:CIvsCO} as dash-dot (red) lines.

\citet{Jiao et al. 2017} gave the linear fitting results of (U)LIRGs: $\log L'_\mathrm{CO(1-0)} = (0.65 \pm 0.02) +  \log L'_\mathrm{[CI](1-0)}$ and $\log L'_\mathrm{CO(1-0)} = (1.19 \pm 0.01) +  \log L'_\mathrm{[CI](2-1)}$ with corresponding correlation coefficients of 0.81 and 0.85, respectively. The correlation coefficients become 0.88 and 0.85 for (U)LIRGs together with Spiderweb Galaxy (for \Cone\ only) and SMGs(for \Ctwo\ only). For comparison, we also present the fitted trends with a fixed slope of 1 of the (U)LIRGs, Spiderweb Galaxy and SMGs as dash-dot (blue) lines in Figure\,\ref{fig:CIvsCO}, and the linear fitting gives: 

\begin{equation}
\log L'_\mathrm{CO(1-0)} = (0.63 \pm 0.02) +  \log L'_\mathrm{[CI](1-0)},
\label{COvsCI10mcfit_LIRG}
\end{equation}
and
\begin{equation}
\log L'_\mathrm{CO(1-0)} =  (1.19 \pm 0.02) +  \log L'_\mathrm{[CI](2-1)},
\label{COvsCI21mcfit_LIRG}
\end{equation}
with scatters of 0.18 dex, and 0.19 dex, respectively.  

Figure\,\ref{fig:CIvsCO} shows that the luminosity of \COone\ correlates tightly and almost linearly with both \Cone\ and \Ctwo\ luminosities for each sub-sample, i.e., the resolved galaxies, and the (U)LIRGs with Spiderweb Galaxy (for \Cone\ only) or SMGs (for \Ctwo\ only), while the two samples spilt up into two distinct linear relations with the intercept of the resolved galaxies increases by $\sim0.3\,$dex, indicating that \CI\ lines can trace the bulk of H$_2$ gas mass in each sub-sample, while neutral carbon abundance and/or carbon excitation temperature might be different in these two sample. Detail analysis will be shown in section\ref{carbon-abundance}.

\subsubsection{\CI\ lines as molecular gas tracers}
\label{Analysis as molecular gas tracers}

\begin{figure*}[!htpb]
   \centering
  \scriptsize 
\centerline{\includegraphics[angle=270, bb= 1 86 430 504, width=0.95\textwidth]{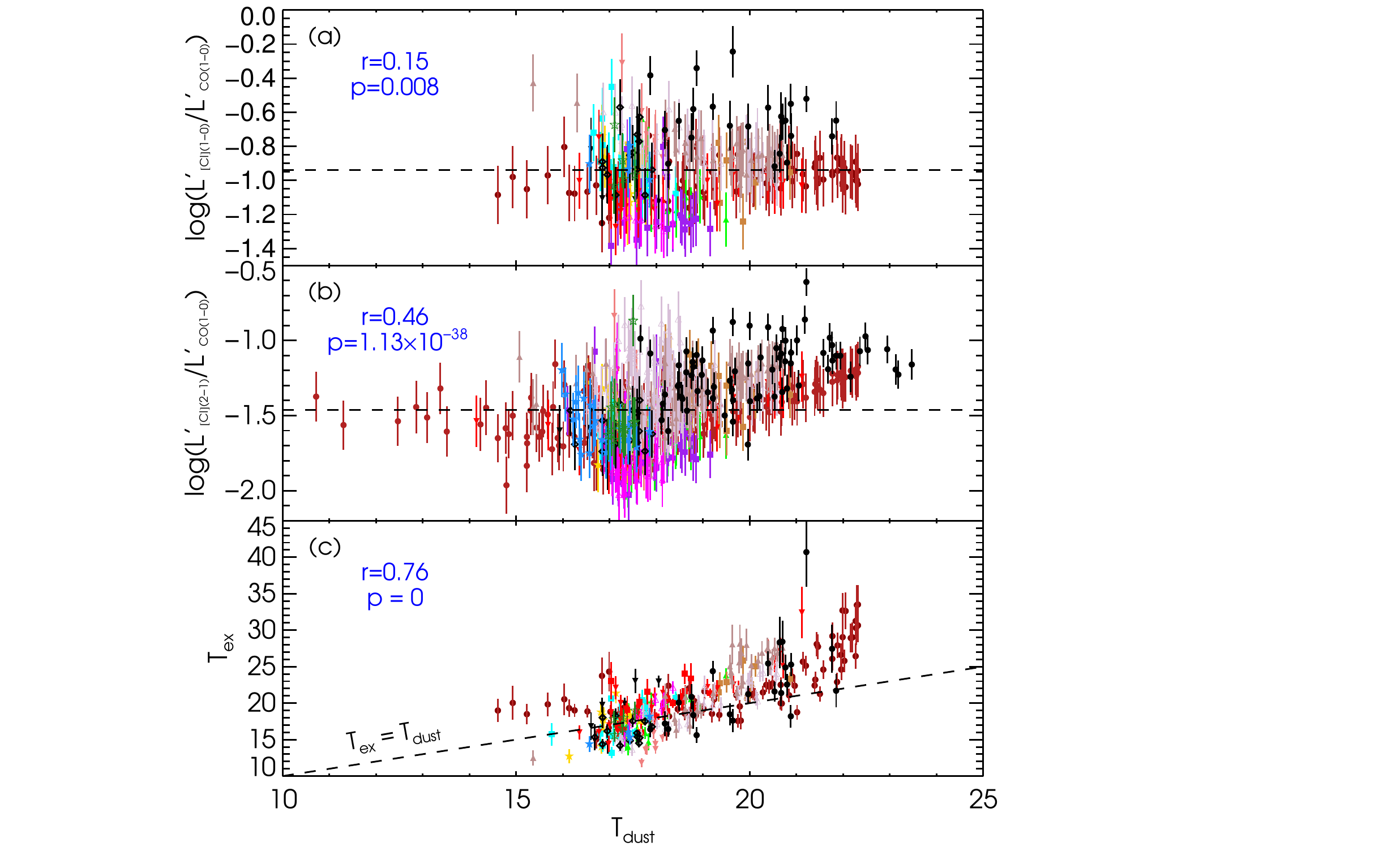}}      
                        
\caption{$L'_\mathrm{[CI](1-0)}$/$L'_\mathrm{CO(1-0)}$ (a), $L'_\mathrm{[CI](2-1)}$/$L'_\mathrm{CO(1-0)}$ (b), and $T_{\rm ex}$ (c) plot against dust temperature which is calculated with color of $f_{70}/f_{160}$ for the resolved galaxies and (U)LIRGs. The meanings of each symbols are same as Figure\,\ref{fig:CIvsCO}. The labelled $r$ and $p$ represent the correlation coefficient and the possibility of no correlation, respectively. The dashed black lines in panels (a) and (b) mark the average ratios of the resolved galaxies and (U)LIRGs.}
\label{fig:color}
\end{figure*}

We have proved that both \CI\ lines can be used as molecular gas tracers in section\ref{CI-CO correlation}. In this section, we further compare the ability of \CI\ lines as molecular tracers to other tracers, such as \COone, and dust.

Figure\,\ref{fig:M82} shows that the ratio of $ L'_\mathrm{[CI](1-0)}$/$L'_\mathrm{CO(1-0)}$ is constant across M~82. Moreover, almost all galaxies in the sample, presented in figure set, regardless of the galaxy types, show constant $ L'_\mathrm{[CI](1-0)}$/$L'_\mathrm{CO(1-0)}$ distributions except some points which may be caused by systematic and/or observational errors. In Figure\,\ref{fig:color}(a), we further present the $ L'_\mathrm{[CI](1-0)}$/$L'_\mathrm{CO(1-0)}$ as a function of the $T_\mathrm{dust}$ for the total resolved sample and (U)LIRGs. The low possible correlation coefficient of r=0.15 and significance of p=0.008 indicating that $ L'_\mathrm{[CI](1-0)}$/$L'_\mathrm{CO(1-0)}$ is constant and likely has no correlations with $T_\mathrm{dust}$. These results prove that the conversion factor of \Cone\ luminosity to molecular gas mass ($X_\mathrm{[CI](1-0)}$) changes with $\alpha_\mathrm{CO}$ proportionally. 
 
We notice that $\alpha_\mathrm{CO}$ not only changes with different galaxy types, but also varies within a galaxy. \citet{Sandstrom et al. 2013} concluded that  $\alpha_\mathrm{CO}$ in the center region of most galaxies shows a factor of $\sim$2 lower value than its galaxy average value, and some can be factors of 5$-$10 below the ``standard" Milky Way value. \citet{Israel 2009} also found that $\alpha_\mathrm{CO}$ in the starburst galaxy center is typically an order of magnitude less than the ``standard" value in the milky way. So the relatively constant ratios of $L'_\mathrm{[CI](1-0)}$/$L'_\mathrm{CO(1-0)}$ within galaxies in Figure\,\ref{fig:M82}, figure set, and Figure\,\ref{fig:color}(a) indicate that $X_\mathrm{[CI](1-0)}$, similar as $\alpha_\mathrm{CO}$, may also changes between different galaxy types and varies within a galaxy. We thus conclude that the $X_\mathrm{[CI](1-0)}$ has consistent ability in tracing H$_2$ gas mass as $\alpha_\mathrm{CO}$ in these galaxies.

However, for high redshift galaxies, the \COone\ is significantly affected by the CMB background \citep{da Cunha et al. 2013, Zhang et al. 2016} and becomes difficult to observe using ground-based telescopes. For a gas kinetic temperatures of 18$\,$K, \citet{da Cunha et al. 2013} found that less than 20$\%$ of intrinsic \COone\ line fluxes can be measured against the CMB at z$>5$. Using a grand design spiral galaxy NGC~628, for which there are high-quality $T_\mathrm{dust}$ \citep{Galametz et al. 2012} as well as a large and fully sampled CO\,(2$-$1) map \citep{Leroy et al. 2009}, \citet{Zhang et al. 2016} concluded that the CO distributions start to diminish for redshifts z$\ge2$, while the \CI\ lines can maintain a larger contrast than the \COone\ against the CMB. Meanwhile, \CI\ lines become accessible for ground mm/submm telescopes for high redshift systems. Moreover, the \CI\ lines are always optically thin even in high column density environments, e.g., star-forming dusty high-z galaxies \citep{Walter et al. 2011, Nesvadba et al. 2019}. The priority of $X_\mathrm{[CI](1-0)}$ becomes obvious in high redshift \citep{Papadopoulos et al. 2004a}.

We also present the $L'_\mathrm{[CI](2-1)}$/$L'_\mathrm{CO(1-0)}$ as a function of $T_\mathrm{dust}$ for the resolved sample and (U)LIRGs in panel b of Figure\,\ref{fig:color}. The possible correlation of r=0.46 and significance of p=$1.13\times10^{-38}$ prove $L'_\mathrm{[CI](2-1)}$/$L'_\mathrm{CO(1-0)}$ increasing with $T_\mathrm{dust}$ moderately. This is consistent with our result in section\ref{Distribution section} that the \Ctwo\ is more sensitive to temperature. \citet{Nesvadba et al. 2019} showed that for a temperature range of 20-50$\,$K, the atomic carbon mass (thus the $\mathrm{H_2}$ mass with a carbon abundance: $X[\mathrm{CI}]/X[\mathrm{H_2}]$) estimated with \Cone\ changes by only about 1$\%$, and the mass estimates from \Ctwo\ changes by more than a factor of 3. \citet{Jiao et al. 2017} also concluded that the $X_\mathrm{[CI](2-1)}$ (the conversion factor of \Ctwo\ luminosity to molecular gas mass) is a worse tracer compare to $X_\mathrm{[CI](1-0)}$ in theory. However, both $X_\mathrm{[CI](1-0)}$ and $X_\mathrm{[CI](2-1)}$ are functions of neutral carbon abundance and $T_{\rm ex}$, and using only one \CI\ line could lead to uncertain estimates of the total molecular gas mass, and best way is to cover both lines. We will further analyze the effects of carbon abundance and temperature on \CI\ tracing ability in section\ref{carbon-abundance}.

In Figure\,\ref{fig:color}(c), we plot the $T_\mathrm{ex}$ with the $T_\mathrm{dust}$ for the  resolved sample and (U)LIRGs, and the labelled possible correlation (r=0.76) and significance (p=0) show that the $T_\mathrm{ex}$ has a positive correlation with dust temperature. As another widely used gas tracer, dust, nevertheless, will be difficult to measure (at low frequencies) because the contrast of the intrinsic dust emission against the CMB decrease dramatically at high redshift \citep[e.g.,][]{da Cunha et al. 2013, Zhang et al. 2016}. However from the $T_\mathrm{ex}$/$T_\mathrm{dust}$ distribution of M~82 in Figure\,\ref{fig:M82}(g), we can see that the \CI\ excitation temperature is higher than the dust temperature in the center and becomes comparable with dust temperature in the disk-region. Meanwhile, most of the data in Figure\,\ref{fig:color}(c) lie above the dashed line ($T_\mathrm{ex}=T_\mathrm{dust}$), indicating that the $T_\mathrm{ex}$ is higher than the $T_\mathrm{dust}$. We further show the histogram-distribution of \CI\ excitation temperatures and dust temperatures for the total resolved galaxies in Figure\,\ref{fig:Tex} with median and mean values labelled on. The uncertainties in the median and mean values represent the MAD and standard deviation, respectively. The histograms also present $T_\mathrm{ex}$ is higher than $T_\mathrm{dust}$. These results prove that \CI\ is easier to be observed in high redshift than (sub-)millimeter dust continuum, and thus maybe a better tracer than dust in distant universe.

\subsection{Neutral carbon abundance}
\label{carbon-abundance}

According to \citet{Weiss et al. 2003, Weiss et al. 2005}, the atomic carbon mass can be derived via:
\begin{eqnarray}
 M_\mathrm{[CI]} & = &C m_\mathrm{[CI]} \frac {8\pi k \nu_0^2}{hc^2A_{\rm 10}} Q(T_{\rm ex}) \frac {1}{3} e^{T_1/T_{\rm ex}}{L'_\mathrm{[CI](1-0)}}\nonumber \\
      &=& 5.706\times 10^{-4}Q(T_{\rm ex})\frac {1}{3} e^{23.6/T_{\rm ex}}{L'_\mathrm{[CI](1-0)}},      
\label{abundanceCI10}    
\end{eqnarray}
using \Cone\ luminosities, and 
\begin{eqnarray}
 M_\mathrm{[CI]} &=&C m_\mathrm{[CI]} \frac {8\pi k \nu_0^2}{hc^2A_{\rm 21}} Q(T_{\rm ex}) \frac {1}{5} e^{T_2/T_{\rm ex}}{L'_\mathrm{[CI](2-1)}} \nonumber \\
     &=&4.566\times 10^{-4}Q(T_{\rm ex})\frac {1}{5} e^{62.5/T_{\rm ex}}{L'_\mathrm{[CI](2-1)}},    
\label{abundanceCI21}     
\end{eqnarray}
with \Ctwo\ luminosities under optically thin and local thermodynamical equilibrium (LTE) assumptions. Among the two equations, $C$ is the conversion between $\mathrm{pc^2}$ to $\mathrm{cm^2}$, and $m_\mathrm{[CI]}$ represents the atomic carbon mass. $A_\mathrm{10}=7.93 \times 10^{-8}\, \mathrm{s^{-1}}$ and $A_\mathrm{21}=2.68 \times 10^{-7}\, \mathrm{s^{-1}}$ are the Einstein coefficients. $T_\mathrm{ex}$ is the \CI\ excitation temperature which can be estimated using equation\\ $T_\mathrm{ex} = 38.8\, \mathrm{K/ln[2.11}/R_\mathrm{[CI]}]$ under optically thin condition \citep{Stutzki et al. 1997}. $Q_{\rm ex}= 1 + 3e^{-T_1/T_{\rm ex}} + 5e^{-T_2/T_{\rm ex}}$ is the \CI\ partition function which depends on excitation temperature $T_{\rm ex}$ with $T_1=23.6\ $K and $T_2=62.5\ $K (the energies above the ground state). Thus carbon mass is a function of $T_{\rm ex}$ and \CI\ luminosities under optically thin and LTE conditions.

\begin{figure}[!htpb]
   \centering
  \scriptsize 
     \includegraphics[width=0.5\textwidth, bb= 50 80 716 515]{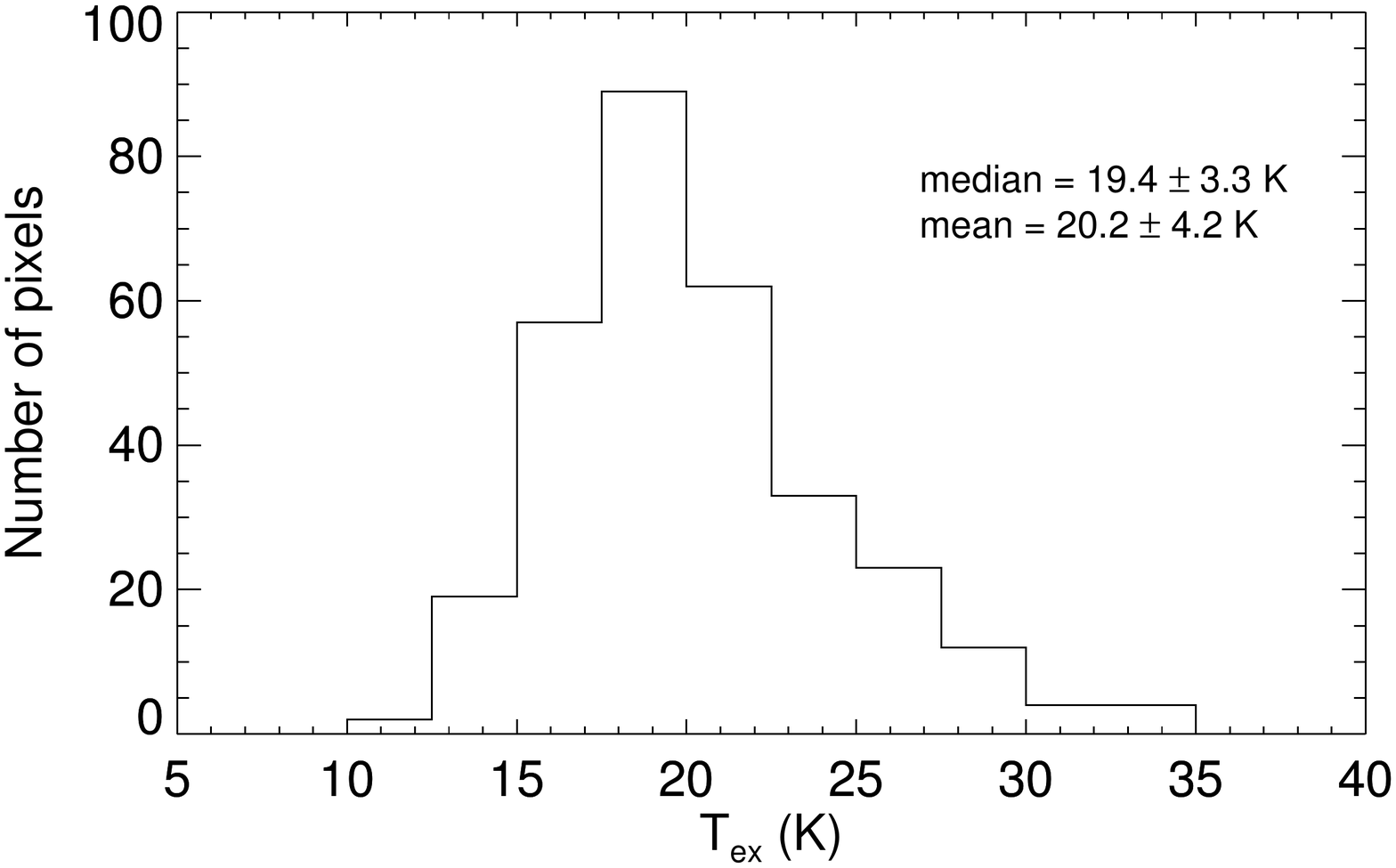}                      
     \includegraphics[width=0.5\textwidth, bb= 50 80 716 515]{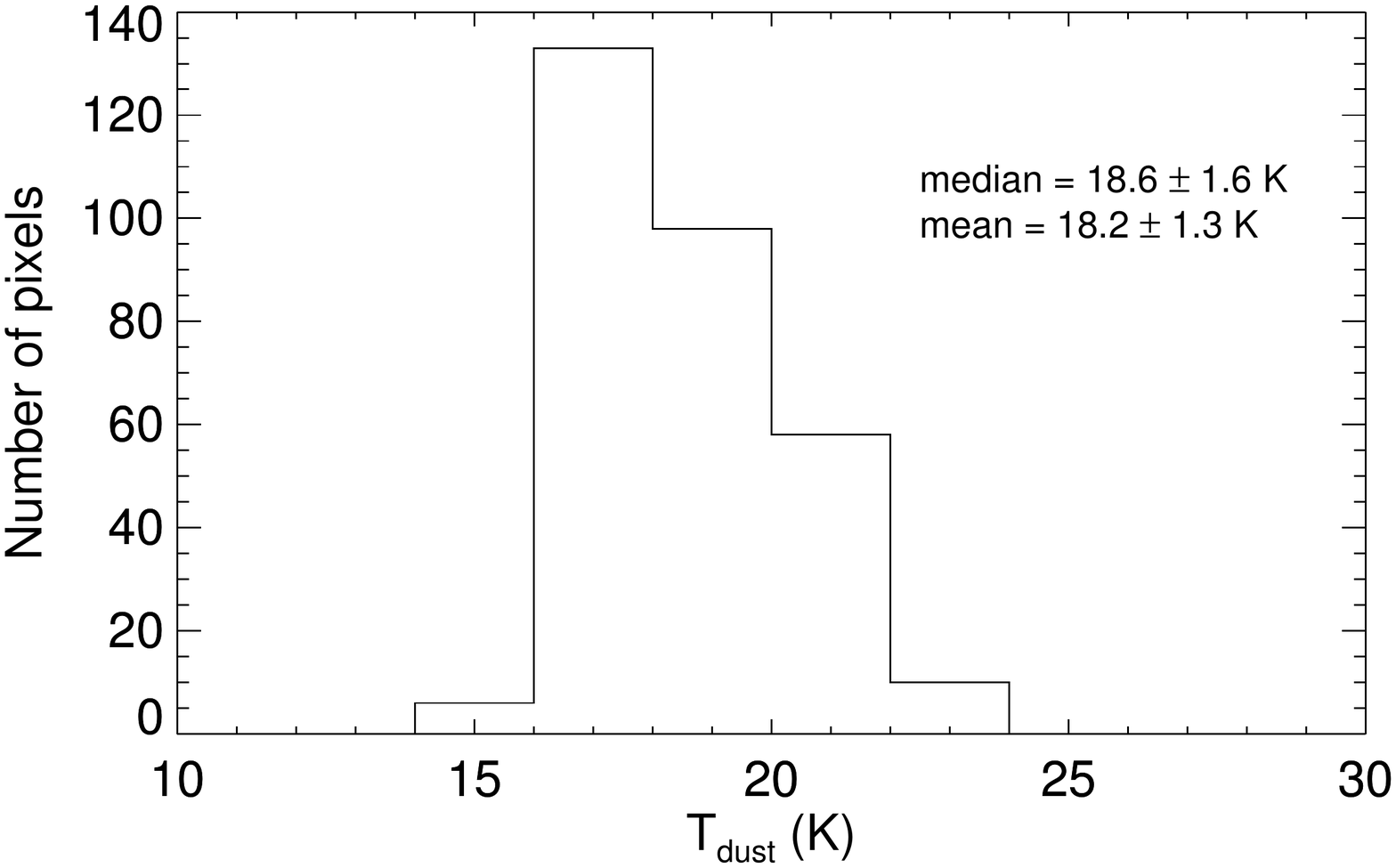}
\caption{Distributions of \CI\ excitation temperatures (top panel) and dust temperatures (bottom panel) for the total resolved samples with median and mean values labelled on.} 
\label{fig:Tex}
\end{figure}

For pixels with both \Cone\ and \Ctwo\ detections, we calculate their \CI\ excitation temperatures using $T_\mathrm{ex} = 38.8\, \mathrm{K/ln[2.11}/R_\mathrm{[CI]}]$. As shown in Figure\,\ref{fig:Tex}, the excitation temperatures mainly concentrate between the range of 11$\,$K to 34$\,$K with a median value of $T_\mathrm{ex}=19.4 \pm 3.3\,$K and a mean value of $T_\mathrm{ex}=20.2 \pm 4.2\,$K. For region with both \CI\ detections of each galaxy, we also calculate the galaxy average $T_\mathrm{ex}$ using the summed \CI\ luminosities, and these excitation temperatures are shown in Table\,\ref{sample}. The galaxy average excitation temperature is in the range of $14 - 26\,$K with an average value of $T_\mathrm{ex} \sim 19.7 \pm 0.5\,$K.

In order to obtain neutral carbon abundance, an independent method is needed to measure H$_2$ mass. For the region with \COone\ and \CI\ detections of each galaxy, we estimate the H$_2$ mass from CO luminosities via $M_\mathrm{H_2} = \alpha_\mathrm{CO}L'_\mathrm{CO}\ M_\odot$. The conversion factor varies within a galaxy \citep[e.g.,][]{Bolatto et al. 2013, Sandstrom et al. 2013}, and we only calculate the global H$_2$ masses and carbon abundance for each galaxy with their galaxy average $\alpha_\mathrm{CO}$. The adopted $\alpha_\mathrm{CO}$ for each galaxy is shown in Table\,\ref{sample}. Particularly, the M~51, M~83, NGC~253, and NGC~1068 are excluded when estimating the carbon abundance due to their uncertain conversion factors. The carbon masses are calculated using equation\,\ref{abundanceCI10} and \ref{abundanceCI21} with \Cone\ and \Ctwo, respectively. Then the global, galaxy-integrated neutral carbon abundance can be estimated using mass ratio between \CI\ and H$_2$: $X[\mathrm{CI}]/X[\mathrm{H_2}]=M(\mathrm{[CI]})/(6M\mathrm{(H_2)})$. The carbon abundance is also calculated for each system using \Cone\ and \Ctwo, and the final values are shown in Table\,\ref{sample}.

The galaxy-wide average carbon abundance is $2.5 \pm 1.0 \times 10^{-5}$. This is comparable  with the usually adopted abundance of $X[\mathrm{CI}]/X[\mathrm{H_2}] \sim 3.0\times 10^{-5}$ \citep{Weiss et al. 2003, Papadopoulos et al. 2004b}, and slightly lower than the result derived by \citet{Alaghband-Zadeh et al. 2013} of $X[\mathrm{CI}]/X[\mathrm{H_2}] = 3.9 \pm 0.4 \times 10^{-5}$ for SMGs. Our results are also lower than the abundance of $X[\mathrm{CI}]/X[\mathrm{H_2}] \approx 8.4 \pm 3.5 \times 10^{-5}$ estimated by \citet{Walter et al. 2011} for SMGs and QSOs at redshift $\sim 2.5$ using CO\,(4$-$3), and slightly higher than the abundance of $1.6 \pm 0.7 \times 10^{-5}$ derived by \citet{Valentino et al. 2018} for the main-sequence galaxies at z $\sim$ 1.2 with CO\,(2$-$1). We also re-calculate the carbon abundance of local (U)LIRGs which have both \CI\ and \COone\ detections in \citet{Jiao et al. 2017} and obtain a mean value of $X[\mathrm{CI}]/X[\mathrm{H_2}] = 8.3 \pm 3.0 \times 10^{-5}$ using \COone\ with a fixed $\alpha_\mathrm{CO} = 0.8$ $M_\odot\  (\mathrm{K\,km\,s^{-1}\,pc^2})^{-1}$. The carbon abundance of (U)LIRGs is $\sim$3 times higher than our resolved sample. \citet{Valentino et al. 2018} found lower carbon abundances in main-sequence galaxies than in high-redshift starbursting systems and SMGs, and concluded that the neutral carbon abundance varies in galaxy types, which is consistent with our results. Using universal carbon abundance may results in biased total molecular gas mass. However, these (U)LIRGs and high-redshift systems all adopted a fixed $\alpha_\mathrm{CO}$ and/or assumed a $T_{\rm ex}$, which may cause deviations in the carbon abundances. On the other hand, our assumptions about $\alpha_\mathrm{CO}$, optically thin and LTE may also result in uncertainties of carbon abundances.

Compared to $\alpha_\mathrm{CO}$ in the sample, which ranges from 1.0 in M~82 to 10.7 in NGC~7331, the calibrated carbon abundance is more constant (ranges from 1.3 $-$ 4.5). The $\alpha_\mathrm{CO}$ of NGC~3521 and NGC~7331 are significantly higher (almost ten times) than other galaxies in our sample. However, both of their carbon abundances are comparable to others. This may indicates that the carbon abundance is more stable than the $\alpha_\mathrm{CO}$ in different environments. We note that the excitation temperature of $T_{\rm ex}$ can be confirmed when both of the \CI\ lines are available, then the carbon abundance is the only fundamental parameter of \CI\ lines to constrain the H$_2$ mass. We thus conclude that the \CI\ lines together may be a better tracer compare to \COone.

However, we need to be careful with the \Cone\ and \Ctwo\ tracers when only one \CI\ line is available. People need to assume an excitation temperature to calculate H$_2$ mass when only one \CI\ line is usable. And the adopted $T_{\rm ex}$ also affects $X_\mathrm{[CI](1-0)}$ and $X_\mathrm{[CI](2-1)}$ differently. \citet{Weiss et al. 2005} showed that the $M_\mathrm{[CI]}$ (thus the H$_2$ mass) estimated from \Cone\ luminosity is sensitive to $T_{\rm ex}$ when the \CI\ excitation temperature is below 20$\,$K (see their Fig.2). And the atomic carbon mass derived with \Cone\ changes by only about 1$\%$ for a temperature range of 20-50$\,$K \citep{Nesvadba et al. 2019}. Therefore, for systems with $T_{\rm ex}$ temperature higher than 20$\,$K, \Cone\ can be a good molecular tracer. For our sample galaxies, as shown in Table\,\ref{sample}, their average $T_{\rm ex}$ is in the range of 14-26$\,$K, and the derived carbon masses change by about 32$\%$. Furthermore \citet{Nesvadba et al. 2019} showed that the carbon mass estimates from \Ctwo\ changes by more than a factor of 3 for a temperature range of 20-50$\,$K. For our sample, the carbon mass changes about a factor of 5 for the temperature range of 14-26$\,$K.

\section{Summary}

We have presented the \Cone\ and \Ctwo\ resolved maps of 15 spiral galaxies which contain one H{\sc ii}, six LINER, three Seyfert, and five starburst galaxies, and compared their \CI\ spatial distributions to \COone\ distribution. For each systems, we have also showed their distributions of \CI-\COone\ luminosity ratios, \RCI, far-infrared color of $f_{70}/f_{160}$, and temperature ratio of $T_\mathrm{ex}$/$T_\mathrm{dust}$. Using statistical method, we have studied the relations of \CI\ luminosities with \COone\ luminosity. We have investigated the dependence of $ L'_\mathrm{[CI](1-0)}$/$L'_\mathrm{CO(1-0)}$ and $L'_\mathrm{[CI](2-1)}$/$L'_\mathrm{CO(1-0)}$, as well as $T_\mathrm{ex}$ on the dust temperature of $T_\mathrm{dust}$. We have also estimated the carbon abundance. Our main findings are as follows:\\

1. Both \Cone\ and \Ctwo\ distribute similarly with  that of \COone\ emission in most of our systems. The $L'_\mathrm{[CI](2-1)}$/$L'_\mathrm{CO(1-0)}$, \RCI, and far-infrared color of $f_{70}/f_{160}$ have centrally-peaked distributions in starbursts, whereas remain relatively constant in LINER galaxies, indicating that active star formation can enhance carbon emissions. \\

2. Both \CI\ luminosities of $L'_\mathrm{[CI]}$ correlate tightly and linearly with $L'_\mathrm{CO(1-0)}$ for the kpc scale points in H{\sc ii}, LINERs, Seyferts, and starbursts: $\log L'_\mathrm{CO(1-0)} = (0.74 \pm 0.12) + (1.04\pm 0.02) \ \log L'_\mathrm{[CI](1-0)}$, and $\log L'_\mathrm{CO(1-0)} = (1.04 \pm 0.08) + (1.08 \pm 0.02) \ \log L'_\mathrm{[CI](2-1)}$, indicating that both \CI\ lines can trace bulk of molecular gas mass in H{\sc ii}, LINERs, Seyferts, and starbursts on sub-kpc scale. Meanwhile, the correlations of $L'_\mathrm{CO(1-0)}$$-$$L'_\mathrm{[CI]}$ in the resolved galaxies, and the (U)LIRGs with Spiderweb Galaxy (for \Cone\ only) or SMGs (for \Ctwo\ only) sample spilt up into two distinct linear relations with $\sim0.3$\,dex offset. \\

3. $L'_\mathrm{[CI](1-0)}$/$L'_\mathrm{CO(1-0)}$ has no correlation with $T_\mathrm{dust}$ and stays constant distribution in most of our systems. $L'_\mathrm{[CI](2-1)}$/$L'_\mathrm{CO(1-0)}$ and $T_\mathrm{ex}$ have weakly and moderately positive correlations with $T_\mathrm{dust}$. We thus conclude that  $X_\mathrm{[CI](1-0)}$ changes with $\alpha_\mathrm{CO}$ proportionally, and $X_\mathrm{[CI](2-1)}$ is sensitive with temperature. The $T_\mathrm{ex}$ is higher than the $T_\mathrm{dust}$ for most of the detected points, indicating that the \CI\ lines are easier to be observed in distance universe than dust.\\

4. Under optically thin and LTE assumptions, the galaxy-wide average \CI\ excitation temperature of the resolved sample is $T_\mathrm{ex} \sim 19.7 \pm 0.5\,$K and carbon abundance is $X[\mathrm{CI}]/X[\mathrm{H_2}] \sim 2.5 \pm 1.0 \times 10^{-5}$, which is lower than the carbon abundance of the (U)LIRGs ($X[\mathrm{CI}]/X[\mathrm{H_2}] \sim 8.3 \pm 3.0 \times 10^{-5}$). The neutral carbon abundance varies in different galaxy types, and using the universal standard carbon abundance may results in biased total molecular gas mass.

\acknowledgments

We thank the anonymous referee for a thorough and helpful report. We thank Emanuele Daddi for the helpful discussions and suggestions. This work is supported by  National Key Basic Research and Development Program of China (Grant No. 2017YFA0402704), National Natural Science Foundation of China (NSFC, Nos. 11673057, 11861131007, 11420101002, and U1531246), and Chinese Academy of Sciences Key Research Program of Frontier Sciences (Grant No. QYZDJ-SSW-SLH008). NL acknowledges support by NSFC grant \#11673028. X.J. acknowledges the support by the NSFC grant 11603075, and Q.-H.T. acknowledges the support by NSFC grant no. 11803090. This publication made use of data from COMING, CO Multi-line Imaging of Nearby Galaxies, a legacy project of the Nobeyama 45-m radio telescope. 

\vspace{5mm}
\software{HIPE \citep{Ott 2010}}

\bibliography{CI_resolved_reference}

\end{document}